\documentclass[]{aa}
\usepackage{graphicx}
\usepackage{txfonts}
\usepackage{mathptmx}
\usepackage{natbib}
\usepackage{url}
\usepackage{longtable}
\usepackage{color}

\usepackage{lscape,rotating} 

\begin{document}

\title{Thermohaline instability and rotation-induced mixing. I - Low- and intermediate-mass solar metallicity stars up to the end of the AGB}

\author{
         Corinne Charbonnel\inst{1,2}\and
          Nad\`ege Lagarde\inst{1}
          }
          
   \institute{
              Geneva Observatory, University of Geneva, Chemin des Maillettes 51, 1290 Versoix, Switzerland \and
              Laboratoire d'Astrophysique de Toulouse-Tarbes, CNRS UMR 5572, Universit\'e de Toulouse, 14, Av. E.Belin, 31400 Toulouse, France\
             }
   \date{Accepted for publication}

\authorrunning{C.Charbonnel \& N.Lagarde}  \titlerunning{Thermohaline instability and rotation-induced mixing}

  \abstract
  % context heading (optional)
   {Numerous spectroscopic observations provide compelling evidence for non-canonical processes that modify the surface abundances of low- and intermediate-mass stars beyond the predictions of standard stellar theory.}
  % aims heading (mandatory)
   {We study the effects of thermohaline instability and rotation-induced mixing in the 1-4~M$_{\odot}$ range at solar metallicity.}
  % methods heading (mandatory)
   {We present evolutionary models by considering both thermohaline and rotation-induced mixing in stellar interior. We  discuss the effects of these processes on the chemical properties of stars from the zero age main sequence up to the end of the second dredge-up on the early-AGB for intermediate-mass stars and up to the AGB tip for low-mass stars.
Model predictions  are compared to observational data for lithium, $^{12}$C/$^{13}$C, [N/C], [Na/Fe], $^{16}$O/$^{17}$O, and $^{16}$O/$^{18}$O in Galactic open clusters and in field stars with well-defined evolutionary status, as well as in planetary nebulae.}
  % results heading (mandatory)
   {Thermohaline mixing simultaneously accounts for the observed behaviour of $^{12}$C/$^{13}$C, [N/C], and lithium in low-mass stars that are more luminous than the RGB bump, and its efficiency is increasing with decreasing initial stellar mass. 
   On the TP-AGB, thermohaline mixing leads to lithium production, although the $^7$Li yields remain negative. Although the $^3$He stellar yields are much reduced thanks to this process, we find that solar-metallicity, low-mass stars remain net $^3$He producers. Rotation-induced mixing is found to change the stellar structure so that  in the mass range between $\sim$ 1.5 and 2.2~M$_{\odot}$ the thermohaline instability occurs earlier on the red giant branch than in non-rotating models. Finally rotation accounts for the observed star-to-star abundance variations at a given evolutionary status, and is necessary to explain the features of CN-processed material in intermediate-mass stars.}
  {Overall, the present models account for the observational constraints very well over the whole mass range presently investigated.}


   \maketitle

%===================================================   Introduction 

\section{Introduction}
	At all stages of their evolution, low- and intermediate-mass stars (LIMS) exhibit the signatures of complex physical processes that require challenging modelling beyond canonical (or standard) stellar theory\footnote{By canonical  we refer to the modelling of non-rotating, non-magnetic stars, in which convection is the only mechanism that drives mixing in stellar interiors.}. The combined effects of rotation-induced mixing, atomic diffusion, and internal gravity waves, have been extensively studied and were shown to account self-consistently for observational patterns of light elements such as lithium and beryllium in main sequence and subgiant LIMS (see e.g. \citealt{ChTa08},  \citealt{Smiljanic09}, and references therein). During the first dredge-up \citep[1DUP;][]{Iben67}, the stellar surface chemical composition is further modified when the deepening convective envelope mixes the external layers with hydrogen-processed material. Dilution then changes the surface abundances of helium-3, lithium, beryllium, boron, carbon, nitrogen, and eventually sodium,  with modification amplitudes that strongly depend on the initial stellar mass and metallicity \citep [e.g.] [] {SwGrRe89, Charbonnel94, BoSa99}. After the completion of the first dredge-up, both standard and recent rotating models \citep{Chaname05,Palacios06} predict no further variations in the surface abundance patterns until the stars start climbing the asymptotic giant branch (AGB). 

Numerous observations provide, however, compelling evidence of a non-canonical mixing process that occurs when low-mass stars reach the so-called bump in the luminosity function on the red giant branch (RGB). At that phase, indeed, the surface carbon isotopic ratio drops, together with the abundances of lithium and carbon, while that of nitrogen increases slightly \citep{Gilroy89,  GiBr91,Luck94,Charbonnel94,Charbonnel98,CharDoNa98, Gratton00,Tautvaisiene00,Tautvaisiene05,Smith02,Shetrone03,Pilachowski03,Geisler05, Spite06,ReLa07,Smiljanic09}. 
Based on lithium observations, \citet{ChBa00} proposed  that intermediate-mass stars suffer from a similar extra-mixing episode when they reach the equivalent of the bump on the early-AGB phase, after helium exhaustion.

Thermohaline mixing has recently been identified as the mechanism that governs the photospheric composition of low-mass bright\footnote{``Bright" refers here to stars that are more evolved than the RGB bump.}  giants \citep[][hereafter CZ07]{ChaZah07a}. 
In such stars, this double diffusive instability is induced by the molecular weight inversion created by the $^3$He($^3$He,2p)$^4$He reaction in the external wing of the hydrogen-burning shell. Indeed  this peculiar reaction converts two particles into three and thus decreases the mean molecular weight, as already pointed out by \citet{Ulrich71} although in a different  stellar context. 
The thermohaline instability is expected to set in 
after the first dredge-up when the star reaches the RGB luminosity bump.
In terms of stellar structure, the RGB bump corresponds to the moment when the hydrogen-burning shell (hereafter HBS) encounters the chemical discontinuity created inside the star by the convective envelope at its maximum extent during the first dredge-up. 
When the source shell (which provides the stellar luminosity on the RGB) reaches 
the border of the H-rich previously mixed zone, 
the corresponding decrease in molecular weight of the H-burning layers induces a drop in the total stellar luminosity, thereby creating a bump in the luminosity function since stars spend a relatively long time at this location \citep[i.e.,][]{Fusipecci90, Charbonnel94, Charbonnel98}. Afterwards H-burning occurs in a region of uniform composition, allowing for the molecular weight inversion due to $^3$He burning to show up and thus enabling the thermohaline instability to set in.
 
Actually it was \citet{Eggleton06} who first drew attention to the destabilizing role of the mu-inversion due to $^3$He-burning in a red giant. 
Their claim was based on a 3D simulation aimed at studying the core helium flash of low-mass red giants with the hydrodynamic code ``Djehuty"  and performed with a mu-profile drawn from a 1D evolutionary sequence \citep{Dearborn06}, which fortuitously demonstrated what the authors called an ``unexpected mixing";  they ascribed it to the well-known Rayleigh-Taylor instability \citep{Eggleton06,Eggleton07}. However CZ07 pointed out that in a star, as the inverse mu-gradient is gradually building up, the first instability to occur and to modify the mu-profile is the double-diffusive instability known in the literature under the generic name of Ç thermohaline instability È \citep{Stern60}. This important precision on the actual nature of the physical process operating in the star was acknowledged by 
\citet[][see also \citealt{DenissenkovPinsonneault08} and \citealt{CantielloLanger2010}]{Eggleton08}; 
%\citet[][although with no reference to CZ07; see also \citealt{DenissenkovPinsonneault08,CantielloLanger2010}]{Eggleton08}; 
it is not just a question of semantics, since these two instabilities - one dynamical and the other double diffusive - proceed on much different timescales. 
It is important to note that thermohaline instability has long been known to develop in other stellar situations whenever inverted molecular-weight gradients are built. This is the case  for instance when helium- or carbon-rich material is deposited at the surface of a star in a mass-transferring binary \citep{StoSim69, Stancliffe07}, when a star accretes heavy elements during planet formation \citep{Vauclair04}, or after the ignition of $^4$He burning in a degenerate shell \citep[i.e., core helium flash; see][]{Thomas67,Thomas70}. It also develops in stars where radiative levitation leads to the accumulation of heavy elements in outer stellar layers \citep{Theado09}. 
Since the term ``thermohaline" is long established in the literature, we see no reason to replace it by the expression ``$\delta \mu$-mixing" as proposed by \citet{Eggleton08}.

CZ07 showed how nicely models including thermohaline mixing as described by the theoretical prescription advocated by \citet[][see \S 2.2]{Ulrich72} do account for the carbon isotopic ratio, as well as for the lithium, carbon, and nitrogen abundances in low-metallicity, low-mass bright giants \citep[see also][]{Stancliffe09}  and simultaneously reduce significantly  the stellar yields of $^3$He with respect to canonical models, as required by measurements of $^3$He/H in Galactic HII regions \citep{Balser94,Balser99a,Bania97,Bania02,Tosi1998, Dearbornetal96,Charbonnel02,Romano03}.
 
Note that the connection between the so-called ``helium-3 problem" and the behaviour of the carbon isotopes in RGB stars was long established.  Indeed although it had not yet been identified, the mechanism responsible for the low $^{12}$C/$^{13}$C values in bright giants was expected to lead to the destruction of $^3$He by a large factor in the bulk of the stellar envelope as initially suggested by \citet[][see also \citealt{Hogan1995,Charbonnel95,Wasserburgetal1995,Weissetal1996,SackBoo99,Eggleton06,Balseretal07}]{Roodetal1984}. 
CZ07 results were confirmed by \citet{Eggleton08} although with a simplistic phenomenological procedure to estimate the diffusion coefficient.
Additionally, \citet{CaLa08,CantielloLanger2010} reported that thermohaline instability can still occur during core helium-burning and beyond in stars that have kept a $^3$He reservoir at these advanced phases. Finally \citet{Stancliffe09} and \citet{Siess09} discussed the impact of this mechanism in low-metallicity intermediate-mass thermally pulsing AGB stars and in super-AGB stars respectively.

The signatures of thermohaline mixing induced by $^3$He-burning have been observed in giants of both open and globular clusters, as well as in field stars including extremely metal-poor giants, and in stars belonging to extragalactic systems such as the Large Magellanic Cloud and the Sculptor galaxy (see references above). 
It appears thus to be a universal process that occurs independently of the stellar environment, although it may be inhibited in some rare RGB stars hosting strong fossil magnetic fields. Indeed \citet{ChaZah07b} examined the effect of a magnetic field on the thermohaline instability, and concluded that in a large fraction of the descendants of Ap stars thermohaline mixing should not occur. The relative number of such stars with respect to non-magnetic objects that undergo thermohaline mixing is very low  \cite[less than 5$\%$, see i.e.,][]{Wolff1968,North1993,Poweretal07} and consistent with the statistical constraint coming from observations of the carbon isotopic ratio in evolved stars as estimated by \citet{CharDoNa98}. It also reconciles the measurements of $^3$He/H in Galactic HII regions with high values of $^3$He observed in a couple of planetary nebulae \citep[for more details see][]{ChaZah07b}.

However, the impact of thermohaline mixing on the stellar chemical properties appears to depend on the initial stellar mass and metallicity, as suggested by the observations of the carbon isotopic ratio (which is the most reliable signature of this mechanism) in giant stars over broad ranges in mass and metallicity (see references above). Until now, the theoretical background for these correlations is not firmly established (e.g., the mass range where thermohaline convection actually modifies the stellar surface composition), and the influence of other quantities such as  the rotation history has never been investigated.   

This is the aim of this series of papers where we investigate the occurrence and the impact of thermohaline mixing in stars of various initial masses and metallicities with non-canonical models, i.e., taking into account self-consistently this mechanism together with rotation-induced mixing.
Here we focus on the solar metallicity case, with the main observational constraints coming from abundance determinations in evolved stars with well defined evolutionary status  up to the early-AGB and belonging to open clusters with turnoff masses between $\sim$ 1 and 4~M$_{\odot}$ or to the field. We also use as observational constraints lithium abundance determinations in low-mass oxygen-rich AGB variables and carbon isotopic ratios in planetary nebulae.
In \S~2 we present the input physics of our models.
In \S~3 we discuss the predictions of our models computed with thermohaline mixing only, and then of our models that also include rotation-induced mixing.
Theoretical predictions are compared to observations in \S~4 before we conclude in \S~5.

%===================================================   Partie 2 INPUT

\section{Input physics for the stellar models}

We present models computed with the code STAREVOL 
\citep {Siess00, Palacios03, Palacios06} at solar metallicity (with Asplund et al.~2005 chemical composition except for Ne for which we use the value derived by \citealt{Cunha06}) for a range of initial stellar masses between 1 and 4~$M_{\odot}$.
In order to quantify precisely the impact of each transport process at the various evolutionary phases, we will show predictions for models computed with the following assumptions: 
(1) Standard models (no mixing mechanism other than convection);
(2) Models including thermohaline mixing only (rotation velocity V=0);
(3) Models including thermohaline mixing and rotation-induced processes  for different initial 
rotation velocities.
All the models are computed up to the end of the second dredge-up on the early-AGB. Some of the low-mass models are followed along the TP-AGB up to the AGB tip.

             \subsection{Microphysics }

We use the OPAL opacity tables \citep{IglRog96} for $T>8000$~K that account for C and O enrichments, and the \citet {Ferguson05} data at lower temperatures. In both cases tables consistent with the assumed initial composition have been generated\footnote{http://opalopacity.llnl.gov 

and http://webs.wichita.edu/physics/opacity}. 
We follow the evolution of 53 chemical species from $^{1}$H to $^{37}$Cl using the nominal NACRE nuclear reaction rates \citep{Argulo99} by default and those 
 given in Appendix A otherwise.
The equation of state is described in details in \citet{Siess00} and accounts for the non-ideal effects due to Coulomb interactions and pressure ionisation. The treatment of convection is based on the classical mixing length formalism with $\alpha_{\rm MLT} = 1.6$, and no convective overshoot is included. The mass loss rates are computed with \citet{Reimers75} formula (with $\eta_R$ = 0.5) up to the early-AGB phase, and with \citet{VaWo93} prescription on the TP-AGB. 

                       \subsection{Thermohaline mixing} 
                                     
The thermohaline instability occurs in a stable stratification that satisfies the Ledoux criterion for convective instability:
\begin{equation}
\nabla_{\rm ad} - \nabla + \left({\varphi \over \delta}\right)\nabla_\mu > 0 ,
\end{equation}
but where the molecular weight decreases with depth: 
\begin{equation}
\nabla_\mu  := {{\rm d} \ln \mu \over {\rm d} \ln {\rm P}} < 0 
\end{equation}
with the classical notations  for $\nabla=(\partial \ln {\rm T}  / \partial \ln {\rm P})$, $\varphi=(\partial \ln \rho / \partial \ln \mu)_{{\rm P,T}}$ and $\delta=-(\partial \ln \rho / \partial \ln {\rm T})_{{\rm P},\mu}$, 
$\nabla_\mu$ and $\nabla_{\rm ad}$ being respectively the molecular weight gradient and  the adiabatic gradient. 

For the turbulent diffusivity produced by the thermohaline instablity we use the prescription advocated by CZ07 based on \citet{Ulrich72} arguments for the aspect ratio $\alpha$ (length/width) of the salt fingers as supported by laboratory experiments \citet{Krish03} and including \citet{Kippen80} extended expression for the case of a non-perfect gas (including radiation pressure, degeneracy):
\begin{equation}
{\rm D_ t} =  {\rm C_ t} \,  {\rm K}  \left({\varphi \over \delta}\right){- \nabla_\mu \over (\nabla_{\rm ad} - \nabla)} \quad \hbox{for} \;  \nabla_\mu < 0, 
\label{dt}
\end{equation}
with K the thermal diffusivity. 
\begin{equation}
{\rm C_ t} = {8 \over 3} \pi^2 \alpha^2 ,  
\end{equation}
and with $\alpha = 5$ (Ulrich 1972) this coefficient is ${\rm C_t}$=658. For consistency reasons we assume actually ${\rm C_t}$=1000 as in CZ07.

                     \subsection {Rotation-induced mixing}
 
For the treatment of rotation-induced mixing we proceed as follows. Solid-body rotation is assumed when the star arrives on the zero age main sequence (ZAMS). Typical initial (i.e., ZAMS) rotation velocities are chosen depending on the stellar mass based on observed rotation distributions in young open clusters \citep{Gaige93}. 
Surface braking by a magnetic torque is applied for stars with an effective temperature on the ZAMS lower than 6900~K that have relatively a thick convective envelope as discussed in \citet {TalCha98} and \citet {ChaTal99}; the adopted braking law follows the description of  \citet{Kawaler88}. 
From the ZAMS on the evolution of the internal angular momentum profile is accounted for with the complete formalism developed by \citet {Zahn92} and \citet {MaeZah98} that takes into account advection by meridional circulation and diffusion by shear turbulence
(for a description of the implementation in STAREVOL, see \citealt {Palacios03}, \citealt{Palacios06}, and \citealt{Decressin09}).
The transport of chemicals resulting from meridional circulation and both horizontal and vertical turbulence is computed as a diffusive process throughout the evolution.
The complete treatment for the transport of angular momentum and chemicals is applied up to the RGB tip or up to the second dredge-up for the stars with initial masses below or above 2.0~M$_{\odot}$ respectively.
The convective envelope is supposed to rotate as a solid body (uniform angular velocity) throughout the evolution; we discuss the implications of this assumption  in \S~3.2.2.
The transport of angular momentum by internal gravity waves (which is efficient only in main sequence stars with effective temperatures on the ZAMS lower than 6500~K, see \citealt {TalCha03}), is neglected. 

In the present work the transport coefficients for chemicals associated to thermohaline and rotation-induced mixings are simply added in the diffusion equation and we do not consider the possible interactions between the two mechanisms, nor with magnetic diffusion. 
As a matter of fact under the present assumptions the thermohaline diffusion coefficient is several orders of magnitude higher than the total diffusion coefficient associated to rotation (see \S3). This is confirmed by \citet{CantielloLanger2010} who also show that magnetic diffusion in RGB stars is much less efficient than thermohaline mixing. 
However, we should keep an eye on future hydrodynamic calculations that are required to evaluate with confidence the possible interactions of thermohaline fingers with differential rotation and magnetic fields in red giants. 

\begin{table}
\caption{Luminosities of the bump (L$_{{\rm b,min}}$ and L$_{{\rm b,max}}$, see Fig.\ref{fig:hrd1p25}), of the evolution point when the thermohaline zone contacts the convective envelope (L$_{\rm c, th}$), and of the RGB tip (L$_{\rm tip}$), for the low-mass stars at various initial rotation velocities.}             
\label{tableL}      
\centering                                                            
	\begin{tabular}{c | c | c | c | c | c}          
	\hline                       
	M & $V_{zams}$& L$_{\rm b,min}$ & L$_{\rm b,max}$ & L$_{\rm c, th}$ &  L$_{\rm tip}$ \\    $(M_{\odot})$ & $(km.s^{-1})$ & (L$_{\odot}$)&  (L$_{\odot}$)& (L$_{\odot})$& (L$_{\odot})$\\
	\hline                                    
	1.25 & 0 & 37 & 45 & 94 & 2821  \\
	        & 50 & 25 & 25 & 64 & 2798  \\                 
	        & 80 & 15 & 16 & 72 & 2807  \\  
	        & 110 &16 & 17 & 72 & 2798 \\  
	\hline 
	1.5 & 0 & 50 & 59 & 784 & 2903  \\
	        & 110 & 27 & 36 & 101 & 2869  \\     
	\hline 
	1.8 & 0 & 74 & 83 & 1907 & 2995  \\
	        & 110 & 51 & 57 & 145 &  2768 \\     
	\hline 	
	2.0 & 0 & 79 & 87 & 1908 & 2994 \\
	        & 110 & 64 & 87 & 250 & 2872  \\     
	        & 180 & 74 & 117 & 256 & 2670  \\  
	        & 250 & 76 & 95 & 232 & 2416  \\  
	\hline 
	\label{table:LbumpLcontact}
		\end{tabular}
\end{table}

\begin{table*}
	\hspace{3cm}
	\caption{Surface values of $^{12}$C/$^{13}$C, [C/Fe], [N/Fe], [Na/Fe], N(Li) and N(Be) at the end of the first and second dredge-up (1DUP and 2DUP respectively) and at the RGB tip (RGB) for models computed under different assumptions: Standard (st, no thermohaline nor rotation-induced mixing), with thermohaline mixing only (th), and with both thermohaline and rotation-induced mixing (th+rot).
	} 
	\scalebox{0.70}{ 
	\centering 
	                 
	\begin{tabular}{c | c | c | c c c | c c c | c c c | c c c | c c c | c c c}      
	\hline\hline                      
	M &  & $V_{ZAMS}$&\multicolumn{3}{c}{$^{12}$C/$^{13}$C}& \multicolumn{3}{c}{[C/Fe]} & \multicolumn{3}{c}{[N/Fe]} & 	\multicolumn{3}{c}{N(Li)} & \multicolumn{3}{c}{N(Be)} & \multicolumn{3}{c}{[Na/Fe]}  \\  
$(M_{\odot})$ & & $(km.s^{-1})$ & 1DUP & RGB & 2DUP & 1DUP & RGB & 2DUP & 1DUP & RGB & 2DUP & 1DUP & RGB & 2DUP & 1DUP & RGB & 2DUP & 1DUP & RGB & 2DUP \\
	\hline\hline                                   
	1 & st & 0 & 28.8 & 28.8 & 25 & -0.05 & -0.05 & -0.07 & 0.16 & 0.16 & 0.21 & 1.1 & 1.1 & 0.85 & 0.34 & 0.34 & 0.15 & 0 & 0 & 0 \\
	   & th  & 0 & 28.8 & 8.1 & 7.7 & -0.05 & -0.09 & -0.10 & 0.16 & 0.25 & 0.28 & 1.1 & -1.3 & -1.87 & 0.33 & -1.65 & -2.1 & 0 & 0 & 0 \\              
	\hline 
	1.1 & th & 0 & 27.35 & 8.9 & 8.5 & -0.07 & -0.11 & -0.12 &  0.22 & 0.28 & 0.33 & 1.34 & -0.58 & -0.97 & 0.27 & -1.33 & -1.63 & 0 & 0 & 0\\
	\hline  
	1.25 & st & 0 & 25.74 & 25.78 & - & -0.10 & -0.10 & - & 0.27 & 0.27 & - &  1.46 & 1.46 & - & 0.18 &  0.18 & - & 0 & 0 & - \\ 
		& th & 0 & 25.6 & 10.4 & 9.9 & -0.10 & -0.12 & -0.14 &  0.37 & 0.31 & 0.33 & 1.45 & 0.07 & -0.26 & 0.18 & -0.97 & -1.23 &  0 & 0 & 0  \\ 
		& th+rot & 50 & 23.69 & 10 & - & -0.12 & -0.14 & - & 0.39 & 0.34 & - &  -4.35 & -3.25 & - & -2.12 & -3.24 &  - & 0 & 0 & -  \\
		& th+rot & 80 & 21.6 & 9.6 & 9.1 & -0.13 & -0.15 & -0.16 & 0.31 & 0.35 & 0.36 & -6.2 & -3.2 & -3.9 & -3.7 & -4.8 &  -5.3 & 0 & 0 & 0 \\
		& th+rot & 110 & 18.6 & 9.1 & 8.5 & -0.14 & -0.16 & -0.18 & 0.33 & 0.37 & 0.38 & -6.87 & -3.25 & -3.94 & -5.12 & -6.21 &  -6.74 &0 & 0 & 0 \\
	\hline 
	1.3 & th & 0 & 24.8 & 11 & - & -0.11 & -0.13 & - & 0.39 & 0.32 & - & 1.46 & 0.22 & - & 0.15 & 0.88 & - & 0 & 0 & -\\
	\hline 
	1.4 & st & 0 & 24.1 & 24.1 & 23.3 & -0.13 & -0.13 & -0.14 & 0.31 & 0.31 & 0.32 & 1.46 & 1.46 & 1.37& 0.01 &  0.01 & 0.03& 0 & 0 & 0\\
	       & th & 0 & 24.2 & 12.14 & - & -0.13 & -0.14 & - & 0.31 & 0.33 & - & 1.46 & 0.46 & - & 0.01 & -0.73 & - & 0 & 0 & -\\
	\hline 
	1.5 & st & 0 & 23.2 & 23.2 & 22.6 & -0.14 & -0.14 & -0.15 & 0.33 & 0.33 & 0.34 &1.45 & 1.45 & 1.37 & 0.06 & 0.06 & 0 & 0 & 0 &  0 \\
	         & th & 0 & 23.3 & 15.3 & - & -0.14 & -0.15 & - & 0.33 & 0.34 & - & 1.45 & 1.03 & - & 0.06 & -0.30 & - & 0 & 0 & -\\ 
		& th+rot & 110 & 21.2 & 12.8 & 11.8 & -0.16 & -0.18 & -0.18 & 0.36 & 0.38 & 0.39 & 0.43 & -0.28 & -0.87  & -0.47 & -1.06 & -1.49 & 0.02 & 0.02 & 0.02\\
	\hline 
	1.7 & st & 0  & 22.5 & 22.5 & 21.9 & -0.16 & -0.16 & -0.17 & 0.36 & 0.36 & 0.37 & 1.41 & 1.41 & 1.34 & -0.02 & -0.02 & -0.07 & 0 & 0 & 0 \\
	       & th & 0 & 22.6 & 18.6 & 16.7 & -0.16 & -0.17 & -0.17 & 0.36 & 0.36 & 0.37 & 1.42 & 1.24 & 0.95& -0.004 & -0.16 & -0.39  & 0 & 0 & 0\\
	\hline 	
	1.8 & th & 0 & 22.2 & 19.9 & - & -0.17 & -0.18 & - & 0.37 & 0.37 & - & 1.41 & 1.32 &  - & -0.03 & -0.11 & - & 0 & 0 & - \\
	      & th+rot & 110 & 20.04 & 15.24 & - & -0.18 & -0.19 & - & 0.31 & 0.31 & - & 0.59 & 0.19 & - & -0.42 & -0.75 & - & 0.15 & 0.15 & -  \\
	\hline 
	1.9 & th & 0 & 21.9 & 20.2 & 18.7 & -0.18 & -0.18 & -0.19 & 0.38 & 0.38 & 0.39 & 1.39 & 1.31 & 1.05 & -0.062 & -0.13 & -0.32 & 0 &0 & 0\\
	\hline 
	2.0 & st & 0  & 21.8 & 21.8 & 21.3 & -0.19 & -0.19 & -0.19 & 0.39 & 0.39 & 0.40 & 1.38 & 1.38 & 1.31 & -0.07 & -0.07 & -0.11 & 0.04 & 0.04 & 0.04\\
	       & th & 0  & 21.8 & 20.8 & 19.6 & -0.19 & -0.19 & -0.20 & 0.40 & 0.39 & 0.40 & 1.37 & 1.31 & 1.10  & -0.07 & -0.12 & -0.28  & 0 & 0 & 0 \\
	       & th+rot & 110 & 19.4 & 16.6 & 16.2 & -0.22 & -0.22 & -0.23 & 0.47 & 0.47 & 0.47 & 0.55 & 0.33 & 0.16 & -0.48 & -0.65 & -0.77 & 0.18 & 0.18 & 0.18 \\
	       & th+rot & 180 & 17.7 & 15.3 & 15 & -0.23 & -0.23 & -0.24 & 0.47 & 0.48 & 0.48  & -0.33 & -0.54 & -0.92  & -0.86 & -1.03 & -1.21 & 0.18 & 0.18 & 0.18\\
	       & th+rot & 250 & 14.8 & 13.4 & 13.2 & -0.22 & -0.23 & -0.23 & 0.47 & 0.48 & 0.48 & -0.96 & -1.15 & -1.3 & -1.29 & -1.44 & -1.55 & 0.22 & 0.22 &0.22\\
	\hline 
           2.5 & st & 0 & 21.2 & 21.2 & 21.1 & -0.21 & -0.21 & -0.21 & 0.45 & 0.45 & 0.45 & 1.34 & 1.34 & 1.25 &  -0.11 & -0.11 & -0.16 & 0 & 0.16& 0.16\\
                  & th & 0 & 21.2 & 21.4 & 20.9 & -0.21 & -0.27 & -0.27 & 0.45 & 0.45 & 0.45 & 1.36 & 1.36 & 1.27 & -0.01 & -0.01 & -0.14 & 0 & 0.16 & 0.16 \\ 
                  & th+rot & 300 & 17.2 & 14.7 & 13 & -0.53 & -0.53 & -0.53 & 0.55 & 0.82 & 0.86 & -4.99 & -4.99 & -5.12 & -3.77 & -3.77 & -3.91 & 0 & 0.76 & 0.85 \\
	\hline 
           2.7  & th+rot & 110 & 19.3 & 19.3 & 19.1& -0.24 & -0.24 & -0.24& 0.52 &0.52 & 0.53 & 0.53 & 0.53 & 0.44 & -0.49 & -0.49 & -0.54& 0.28 & 0.28 & 0.28\\
		& th+rot & 250 & 16.4 & 16.4 & - & -0.28 & -0.28&  - & 0.56 &0.56 & - &  -0.95 &  -0.95 & - & -1.28 & -1.28 & - & 0.32 & 0.32 & -\\
		& th+rot & 300 & 18 & 15.1 & - & -0.53 & -0.53 & - & 0.83 & 0.84 & - & -4.9 & -4.9 & - & -3.79 & -3.79 &  - & 0.76 & 0.76 & - \\
	\hline 
	3.0  & th & 0 & 20.85 & 20.84 & 20.55 & -0.21 & -0.21 & -0.22 & 0.47 & 0.48 & 0.48 & 1.30 & 1.30 & 1.15& -0.15 & -0.15 & -0.23 & 0.031 & 0.22 & 0.23\\
	\hline
         4.0 & st & 0 & 20.48 & 20.48 & 19.94 & -0.21 & -0.21& -0.22 & 0.49 &0.49 & 0.50 & 1.28 & 1.28 & 1.02 & -0.16 & -0.16 & -0.33& 0 & 0.26 & 0.28\\
         	      & th & 0 & 20.49 &20.49 & 19.86 & -0.20& -0.20& -0.22& 0.49 & 0.49 & 0.51 & 1.29 & 1.29 & 1.03 & -0.16 &-0.16 &-0.33& 0.25 & 0.25 & 0.28\\
               & th+rot & 300 & 14 & 14 & 13.82&  -0.31 & -0.31 & -0.32 & 0.61 & 0.61 & 0.62 & -1.76 & -1.77 & -1.94 & -1.77 & -1.77 &  -1.87 & 0.4 & 0.4 & 0.42 \\ 
         \hline                
	\end{tabular}}
	\label{tablesurfabundances}

\end{table*}

%====================================================================== Partie 3 Modeles theoriques 

\section{Theoretical predictions}

We first consider the case of low-mass stars that ignite helium-burning by a flash at the tip of the RGB well above (in terms of luminosity) the bump. With the considered metallicity and input physics this corresponds to stars with initial masses below $\sim$2.2~M$_{\odot}$. We present detailed predictions for a 1.25~M$_{\odot}$ model computed without and with rotation (but with thermohaline mixing in both cases) in \S~3.1 and 3.2 respectively, and discuss the uncertainties on the thermohaline diffusivity in \S~3.3. \S~3.4 is devoted to the case of stars in the mass range 1.5-2.2~M$_{\odot}$.
Then in \S~3.5 we shortly discuss the predictions for intermediate-mass stars.

\subsection{1.25~M$_{\odot}$ model with thermohaline mixing only}

\begin{figure}
	\centering
		\includegraphics[angle=0,width=9cm]{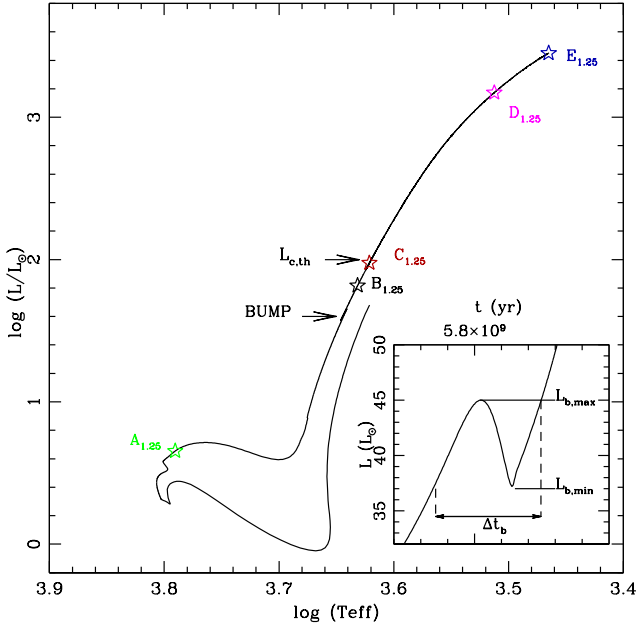}
		 \caption{Evolutionary track (from the pre-main sequence up to the tip of the RGB) of the 1.25~M$_{\odot}$ model computed with thermohaline mixing only. The minimum and maximum luminosity of the bump are indicated (L$_{\rm b, min}$ and  L$_{\rm b, max}$ respectively) as well as the luminosity L$_{\rm c, th}$ at which the thermohaline instability reaches the bottom of the convective envelope. Open symbols labelled A$_{1.25}$ to E$_{1.25}$ correspond to evolution points for which some stellar properties are discussed in the text. The panel inserted on the right of the figure shows the evolution of the stellar luminosity around the bump as a function of time. $\Delta$t$_{\rm b}$ is the time spend by the star within the luminosity bump and equals to 3.9$\times 10^7$ yrs in the present case.}
	\label{fig:hrd1p25}
\end{figure}

Figure~\ref{fig:hrd1p25} presents the evolutionary track in the Hertzsprung-Russell diagram (HRD) of the 1.25~M$_{\odot}$ model computed with thermohaline mixing only (no rotation). Several points are selected along the track in order to discuss the evolution of some relevant stellar properties. A$_{1.25}$ corresponds to the turnoff, when the hydrogen mass fraction in the stellar core is below $10^{-8}$. B$_{1.25}$ is chosen at intermediate luminosity between the bump (which minimum and maximum luminosities, L$_{\rm b, min}$ and  L$_{\rm b, max}$, are also shown) and the moment when the thermohaline zone ``contacts" the convective envelope (see below). C$_{1.25}$ stands at the ``contact" luminosity L$_{\rm c, th}$ where surface abundances start changing due to thermohaline mixing.  D$_{1.25}$ and E$_{1.25}$ are close to and at the tip of the RGB (then the mass of the helium core is 0.428 and 0.486~M$_{\odot}$ respectively, for a total stellar mass of 1.14 and 1.03~M$_{\odot}$).

\subsubsection{Main sequence and subgiant branch}

\begin{figure}
	\centering
		\includegraphics[angle=0,width=9cm]{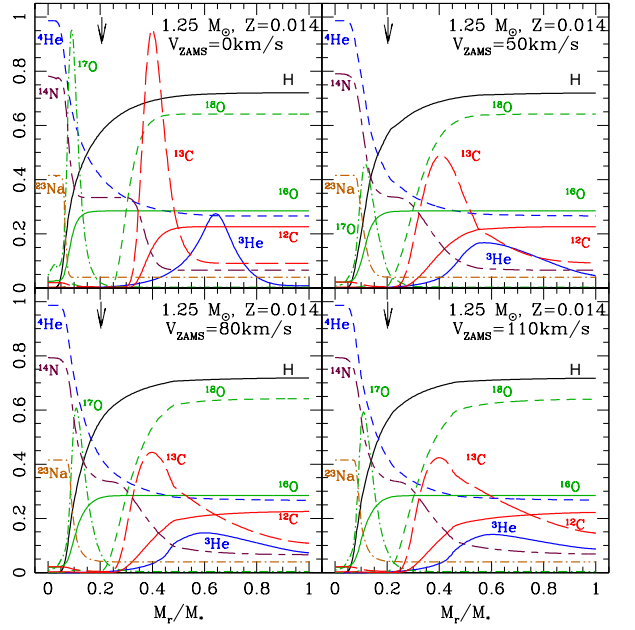}
	  \caption{Chemical structure at the turnoff of the 1.25~M$_{\odot}$ star computed for different initial rotation velocities as indicated. The mass fractions are multiplied by 100 for $^3$He, $^{12}$C, and $^{14}$N, by 2500 for $^{13}$C, by 50, 900, and 5$\times 10^4$ for $^{16}$O, $^{17}$O, and $^{18}$O respectively, and by 1500 for $^{23}$Na.
	  The vertical arrows show, in all cases, the maximum depth reached by the convective envelope at its maximum extent during the first dredge-up.}
	\label{fig:profil_1.25_diffrot_Tof}
\end{figure}

Figure~\ref{fig:profil_1.25_diffrot_Tof} depicts the chemical structure of a 1.25~M$_{\odot}$ star at the end of central hydrogen-burning (point A$_{1.25}$; top left panel for the present case without rotation-induced mixing). The most fragile elements lithium, beryllium, and boron, which burn at relatively low-temperatures and are preserved only in the most external stellar layers, are not shown here. On the pre-main sequence, pristine deuterium is converted to $^3$He, while on the main sequence H-burning through the pp-chains builds up a $^3$He peak at M$_{\rm r}$/M$_* \sim$0.65. Deeper inside the star the $^{13}$C peak  results from the competition between the $^{12}$C(p,$\gamma)^{13}$N($\beta +\nu)^{13}$C and $^{13}$C(p,$\gamma)^{14}$N reactions.  
 $^{12}$C and $^{16}$O are partially converted into $^{14}$N which abundance profile presents a double plateau.  ON-cycling results in $^{18}$O depletion and in the building up of a $^{17}$O peak. In the very central regions, $^{23}$Na is produced through proton capture by $^{22}$Ne.

When the star moves towards the RGB its convective envelope deepens and engulfes most of the regions that have been nuclearly processed (1DUP). In Fig.~\ref{fig:profil_1.25_diffrot_Tof}  the maximum depth reached by the convective envelope is indicated by the vertical arrow. The so-called first dredge-up results in severe changes in the surface chemical properties of the star (see Fig.~\ref{fig:surfaceabundances1p25}) 
since surface material is diluted with matter enriched in $^3$He, $^{13}$C, and $^{14}$N, but depleted in $^{12}$C and $^{18}$O. In the standard models the surface abundances are not predicted  to change anymore after the end of the first dredge-up until the star reaches the early-AGB. However, as we shall see below, thermohaline mixing induces a second modification of the stellar chemical appearance on the upper end of the RGB.

\subsubsection{Predictions up to the RGB tip}

\begin{figure}
	\centering
		\includegraphics[angle=0,width=7.8cm,clip=true,trim= 0cm 0cm 0cm 1cm]{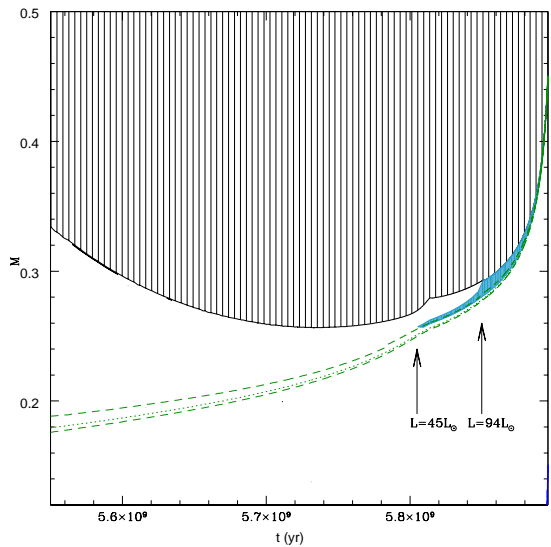}
	  \caption{Kippenhahn diagram for the 1.25~M$_{\odot}$ star computed with thermohaline mixing (no rotation-induced processes). Here we focus on the evolution phase around the RGB luminosity bump (located at a total stellar luminosity of $\sim$ 45~L$_{\odot}$ as indicated by the left arrow). Hatched area is the convective envelope. Green dotted lines delimit the hydrogen-burning shell above the degenerate helium core. The zone where the thermohaline instability develops is shown in hatched blue. The right arrow indicates the total stellar luminosity (L$_{\rm c, th} \sim$ 94~L$_{\odot}$) at which the thermohaline region extends up to the convective envelope and thus when the surface chemical composition starts changing.}
	\label{fig:kippen_1p25_th_avL}
\end{figure}

As discussed in \S~1, the thermohaline instability induced by the $^3$He($^3$He,2p)$^4$He reaction is able to set in on the RGB only after the star has reached the luminosity bump, when the HBS crosses the molecular weight barrier left behind by the convective envelope when it reached its maximum extent. In the case where thermohaline mixing is the only transport process considered within radiative regions (i.e., rotation-induced mixing being neglected), we find that for solar metallicity stars with initial mass lower or equal to 1.5~M$_{\odot}$, the thermohaline instability soon extends 
between the external wing of the HBS and the base of the convective envelope (see Table~1). 
This is shown for the 1.25~M$_{\odot}$ model in Fig.~\ref{fig:kippen_1p25_th_avL} where the thermohaline region is indicated in blue. For this model, the bump is located between L$_{\rm b, min}$=37~L$_{\odot}$ and L$_{\rm b, max}$=45~L$_{\odot}$, and the thermohaline instability contacts the convective envelope when the stellar luminosity is L$_{\rm c, th} \sim$ 94~L$_{\odot}$. As far as timescales are concerned, the 1.25~M$_{\odot}$ model spends 3.9$\times 10^7$ years in the bump (i.e., in the luminosity interval between L$_{\rm b, min}$ and  L$_{\rm b, max}$, see Fig.~\ref{fig:hrd1p25}), and then reaches L$_{\rm c, th}$ after 2.65.10$^7$ years. 

\begin{figure}
	\centering
		\includegraphics[angle=0,width=9cm]{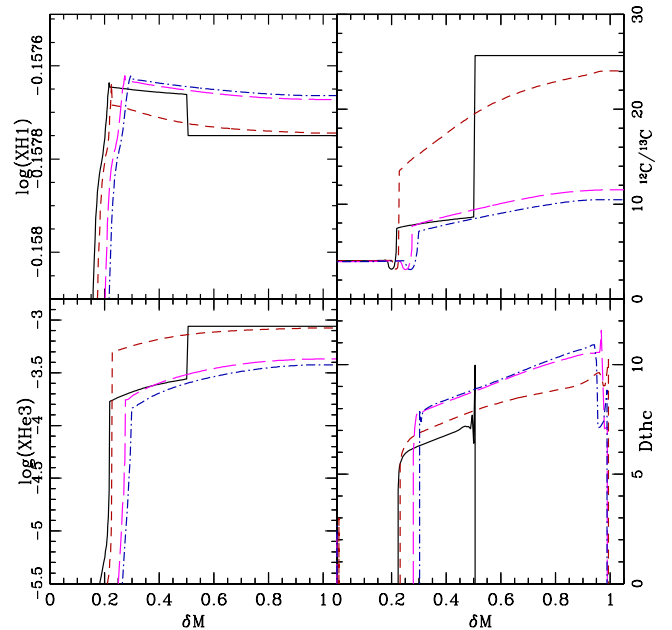}
		\caption{Profiles of the abundances (in mass fraction) of H, $^3$He, of the carbon isotopic ratio, and of the thermohaline diffusion coefficient at various evolution points on the RGB (see Fig.~\ref{fig:hrd1p25}): Slightly above the bump (point B$_{1.25}$, black solid line), at the luminosity of the contact (point C$_{1.25}$, red dashed line), close from the RGB tip (point D$_{1.25}$, magenta long dashed line), and at the RGB tip (point E$_{1.25}$, blue dot-dashed line). The abscissa is the scaled mass coordinate $\delta$M that allows a blow up of the region of interest ($\delta$M=0 at the base of the HBS and $\delta$M=1 at the base of the convective envelope).}
		\label{fig:Dthc_X_1p25}
\end{figure}

\begin{figure}
	\centering
		\includegraphics[angle=0,width=9cm]{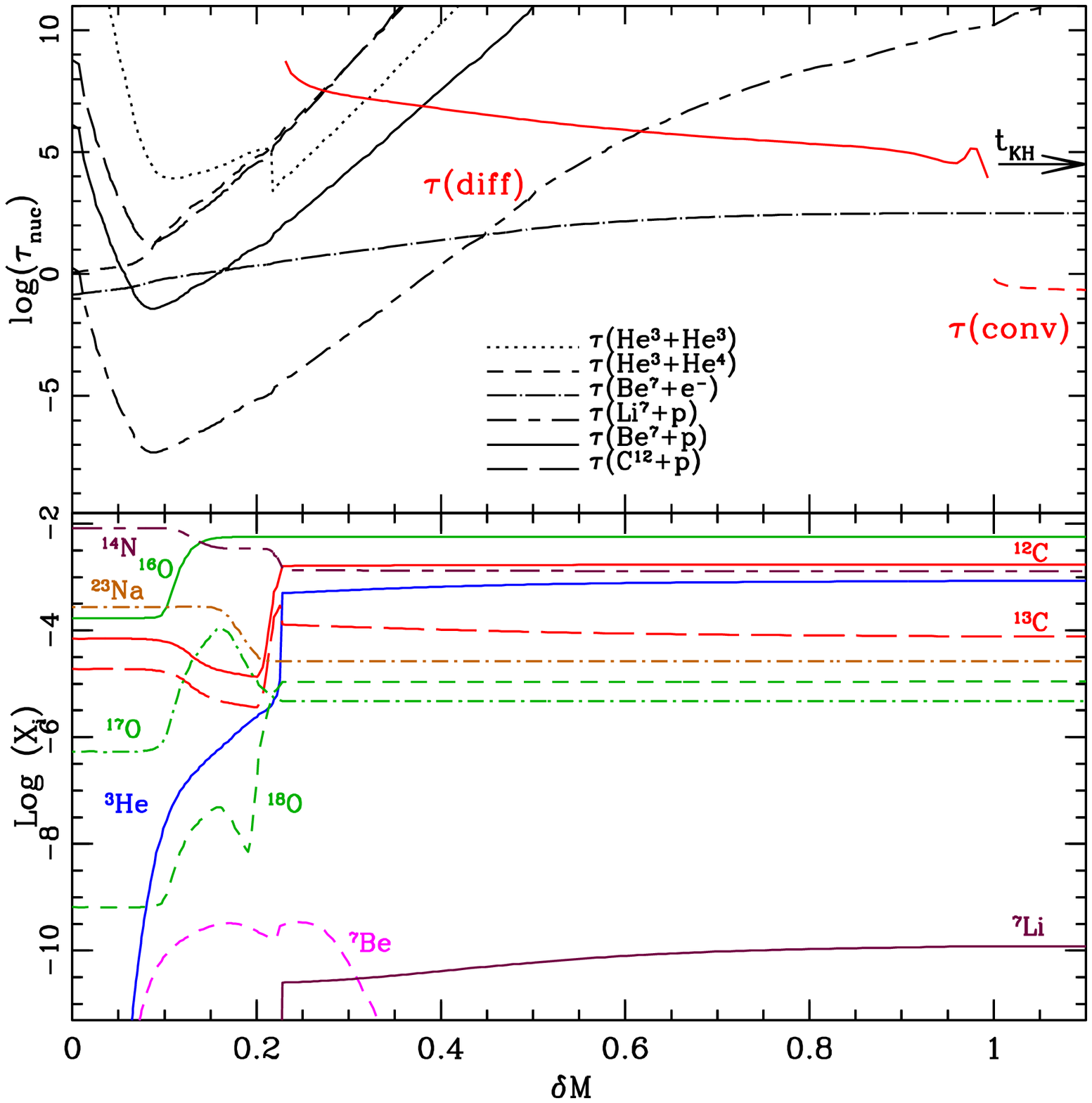}
		\caption{1.25~M$_{\odot}$ model with thermohaline mixing (no rotation) at the evolution point C$_{1.25}$ -- {\it (top)} Lifetime of $^3$He, $^7$Li, $^7$Be, and $^{12}$C, and convective and thermohaline diffusive timescales (in years). The transition between the convective envelope and the radiative region occurs at $\delta$M=1.  {\it (bottom)} Abundance profiles (in mass fraction) of relevant species in the same region.}
		\label{fig:tnuc_Xi_1p25_th_v0_aucontact}
\end{figure}

Figure~\ref{fig:Dthc_X_1p25} depicts the temporal evolution along the RGB 
of the abundance profiles of H, $^3$He, of the carbon isotopic ratio, and of the thermohaline diffusion coefficient D$_{\rm thc}$. The variable $\delta$M is a relative mass coordinate allowing for a blow-up of the radiative region above the HBS. 

Namely, 
$$\delta {\rm M} = \frac{M_r - M_{\rm HBS}}{M_{\rm BCE} - M_{\rm HBS}}.$$

It is equal to 1 at the base of the convective envelope M$_{\rm BCE}$ and to 0 at
the base of the HBS M$_{\rm HBS}$ (which is defined as the depth where the hydrogen mass fraction equals $10^{-10}$) . 
On each graph the black solid lines correspond to the evolution point B$_{1.25}$ on the HRD when the thermohaline region is still quenched in a very tiny region, while the other curves correspond to latter times (points C$_{1.25}$, D$_{1.25}$, and E$_{1.25}$) when the thermohaline instability has extended up to the base of the convective envelope. 
The maximum depth of the thermohaline unstable region corresponds to the layer where  the $^3$He($^4$He,$\gamma$)$^7$Be reaction becomes faster than $^3$He($^3$He,2p)$^4$He ($\delta$M$\sim$0.2; see upper panel in 
Fig.~\ref{fig:tnuc_Xi_1p25_th_v0_aucontact}). There the hydrogen profile shows a peak which maximum is located at the depth where the reaction $^3$He($^3$He,2p)$^4$He is the fastest.  
As soon as the thermohaline instability sets in, fresh protons diffuse outwards, spreading out the molecular-weight inversion and enlarging the thermohaline region until it reaches the convective envelope. Simultaneously $^3$He diffuses from the convective envelope inwards, which fuels the thermohaline instability. $^{12}$C and $^{13}$C diffuse respectively inwards and outwards, leading to a decrease of the surface carbon isotopic ratio.  $^{14}$N also diffuses outwards. Among the oxygen isotopes, only $^{18}$O is affected, which leads to a slight increase of the $^{16}$O/$^{18}$O surface ratio while $^{16}$O/$^{17}$O does not change. 
Elements with higher atomic numbers  such as $^{23}$Na, which burn or are produced at higher temperature than that of the maximum depth of the thermohaline region, are unaffected. 
On Fig.\ref{fig:tnuc_Xi_1p25_th_v0_aucontact} one can see also that in the external wing of the HBS below $\delta {\rm M} \sim0.6$, the thermohaline diffusion timescale $\tau$(diff) is longer than both $\tau (^7$Be + e$^-$) and $\tau (^7$Li+p). 
As a consequence no fresh $^7$Li shows up at the stellar surface on the RGB for  the value of the coefficient C$_{\rm t}$ chosen for the computations. Rather, the surface $^7$Li abundance decreases as
this fragile element is drained from the convective envelope downwards. For the 1.25~M$_{\odot}$ non-rotating model the value of N(Li)\footnote{N(Li)=log[n(Li)/n(H)]+12} at the RGB tip is $\sim 0$ (see Table 2 and Fig.~\ref{fig:Li_1p25_V0_Ct1e3_Ct1e4} in \S~3.3).

\begin{figure}
	\centering
		\includegraphics[angle=0,width=9cm]{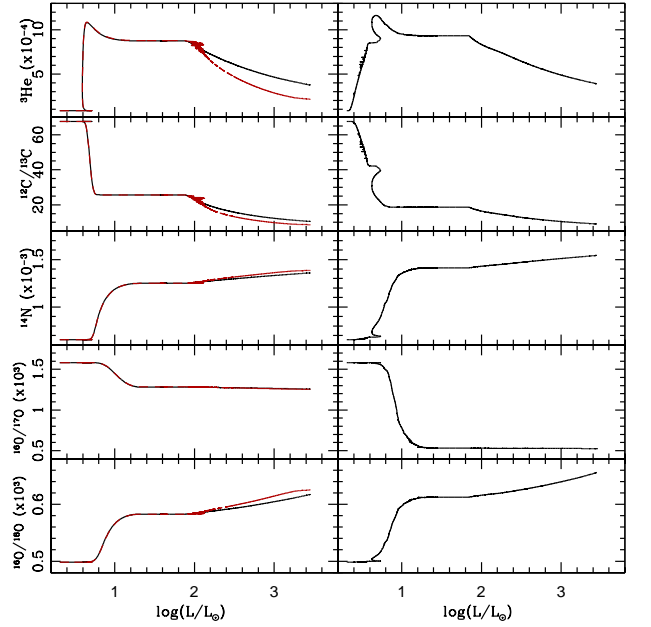}
	  \caption{Evolution of the surface abundances of $^3$He and $^{14}$N (in mass fraction) and of the carbon and oxygen isotopic ratios as a function of stellar luminosity for the 1.25~M$_{\odot}$ models computed without {\it (left)} and with rotation (initial rotation velocity of 110~km.sec$^{-1}$, {\it right}). Thermohaline instablity is accounted for in both cases. In the left panels the black and red lines correspond to C$_{\rm t} = 10^3$ and $10^4$ respectively, while in the right panels C$_{\rm t} = 10^3$. Predictions are shown from the zero age main sequence up to the RGB tip. 
	  In the rotating case the first dredge-up starts at lower luminosity and ends at log(L/L$_{\odot}$)$\sim$1 with slightly stronger abundance modifications than in the standard case. Also, thermohaline mixing sets in slightly earlier on the RGB at log(L$_{\rm c,th}$/L$_{\odot}$)$\sim$1.9 when rotation is accounted for.
	  }
	\label{fig:surfaceabundances1p25}
\end{figure}

The evolution of the surface abundances from the zero age main sequence up to the RGB tip 
 is shown on Fig.\ref{fig:surfaceabundances1p25} (left panels for the present case without rotation-induced mixing).
 One sees clearly the signatures of the first dredge-up at Log(L/L$_{\odot}) \sim$ 0.6 that leads to an increase of the surface abundances of $^3$He, $^{13}$C, $^{14}$N, and $^{17}$O, and of the $^{16}$O/$^{18}$O ratio, and to a decrease of $^{12}$C, $^{18}$O, of the $^{12}$C/$^{13}$C and  $^{16}$O/$^{17}$O ratios  as the convective envelope digs into the regions that have been nuclearly processed on the main sequence (see Fig.\ref{fig:profil_1.25_diffrot_Tof}). 
 Then all surface abundances level off while the star ascends the RGB up to the bump\footnote{In the standard case without thermohaline mixing, the abundances obtained at the end of the first dredge-up remain unchanged until the convective envelope deepens for the second time on the early-AGB.}. At log(L/L$_{\odot}$)$\sim$2, thermohaline mixing does lead to a second change in the stellar surface composition. 
It induces in particular a second drop of the carbon isotopic ratio and efficiently destroys $^3$He  and $^7$Li (see Fig.~\ref{fig:Li_1p25_V0_Ct1e3_Ct1e4}) while $^{14}$N and  $^{16}$O/$^{18}$O slightly increase.
We note that the $^3$He surface abundance at the tip of the RGB is much reduced compared to its value after the first dredge-up, although it remains higher than the initial value the star is born with. This is due to the combination of several factors related to the thermohaline diffusion timescale on one hand, and to the compactness of the HBS. More specifically, the final $^3$He surface abundance can in principle decrease down to the value of $^3$He at the bottom of the thermohaline unstable region (i.e., at $\delta$M$\sim$0.3 in the 1.25~M$_{\odot}$ case; see Fig.~\ref{fig:Dthc_X_1p25}). However, the timescale for thermohaline diffusion in this model is such that the surface $^3$He at the RGB tip saturates at a higher value.
In the case of low-mass, low-metallicity stars presented in CZ07, the thermohaline unstable region is more compact and has a steeper temperature gradient, 
resulting in a more efficient decrease of the surface $^3$He abundance than in the present case. 
The same reasoning applies to the surface abundance changes in $^{12}$C and $^{13}$C so that the carbon isotopic ratio at the tip of the RGB saturates here to a value of the order of 10, while lower ratios closer from the equilibrium value are reached in low-mass, low-metallicity stars (see Fig.~3 and 4 of CZ07).

\subsubsection{Predictions on the AGB}

\begin{figure}
	\centering
		\includegraphics[angle=0,width=9cm]{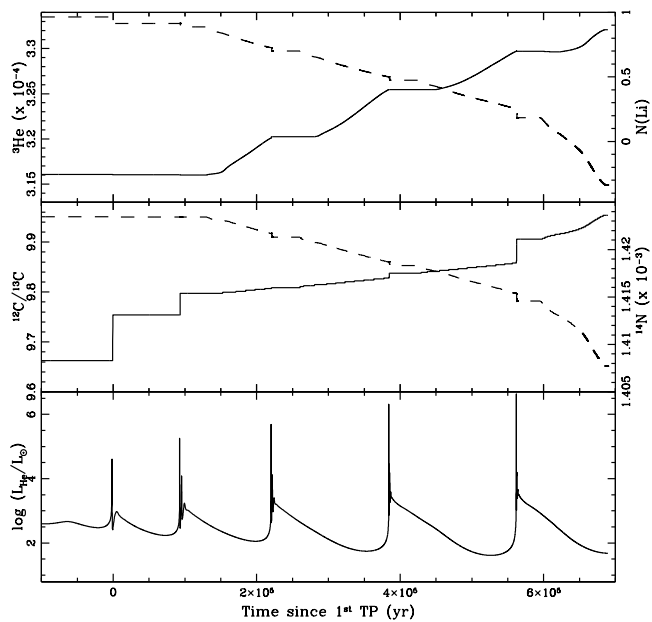}
			  \caption{Evolution of the surface abundances of  $^3$He and $^7$Li ({\it top}, dashed and solid lines respectively), of $^{14}$N and of the carbon isotopic ratio ({\it middle}, solid and dashed lines respectively), and of the helium-burning luminosity and total stellar mass ({\it bottom}, solid and dashed lines respectively) on the TP-AGB of the 1.25~M$_{\odot}$ model computed with thermohaline mixing only. The abscissa is the time since the first thermal pulse.}
	\label{fig:TPs_1p25_v0}
\end{figure}

After helium ignition at the RGB tip, the total stellar luminosity drops down to the location of the clump (at Log(L/L$_{\odot}) \sim 1.7$ for the 1.25~M$_{\odot}$ model discussed here). 
A very modest decrease of the surface $^7$Li and $^{12}$C/$^{13}$C occurs when the star starts ascending the early-AGB due to a second deepening of the convective envelope (see Table 2). Also,  $^3$He (in mass fraction) decreases very slightly from 3.7.10$^{-4}$ to 3.3.10$^{-4}$.
In other words, some $^3$He remains in the external convective layers when the star enters the thermal pulse phase on the AGB (TP-AGB) and is available to drive thermohaline mixing further as discussed below. 

This 1.25~M$_{\odot}$ model was computed until the end of the superwind phase (at that point the total stellar mass is 0.567M$_{\odot}$ and the mass of the envelope is 0.017M$_{\odot}$). Meanwhile it has undergone five thermal pulses. 
In the present study we assume no convectively induced extra-mixing below the convective envelope as usually required to induce the third dredge-up in low-mass TP-AGB models 
\citep{Herwigetal97, Herwig00, Herwigetal07, Mowlavi99,Karakas02,Karakas10}.
Consequently we do not expect our models to mimic carbon-rich stars (but see \S~3.4).
Such a study is out of the scope of the present work and TP-AGB predictions based on the present computations but including also different overshoot prescriptions will be presented in a future paper.

Let us note, however, that we find that with the present assumptions thermohaline mixing is active during each interpulse and modifies the surface abundances as can be seen on Fig.\ref{fig:TPs_1p25_v0}. As on the RGB, thermohaline mixing changes, but in a very modest way, the abundances of $^3$He, $^{12}$C, $^{13}$C, $^{14}$N, $^{17}$O, and $^{18}$O,  while  heavier elements are unaffected. 

Also and in contrast with what happened on the RGB, the surface abundance of $^7$Li now increases (see Table~\ref{tableNLiTPAGB}). 
This is due to the mixing efficiency that allows the transport of fresh $^7$Be outwards to regions cool enough for $^7$Li to survive. In the present case the production of $^7$Li is significant, and the surface N(Li) reaches a value of $\sim$0.9 (then the star has a total luminosity Log(L/L$_{\odot}$) between $\sim$3.0 and 3.6). We thus confirm, but this time at solar metallicity, the finding by \citet{Stancliffe10} that thermohaline mixing does increase the surface Li abundance in low-mass TP-AGB stars. Whether these Li-rich objects can simultaneously be carbon-rich will require further investigation of the TP-AGB phase including parametric convective overshoot as mentioned before.
As can be seen in Table~\ref{tableNLiTPAGB}, the total stellar yields for lithium remain negative.

\subsection{1.25~M$_{\odot}$ model with thermohaline and rotation-induced mixing}
Let us now discuss the case of a 1.25~M$_{\odot}$ model that does take into account both thermohaline instability and rotation-induced mixing as described in \S~2.2 and 2.3.

\subsubsection{Main sequence and subgiant branch}

As described above,  the thermohaline instability induced by $^3$He-burning sets in only on the RGB after the bump. 
Before a star reaches that phase, the modifications of its internal and surface chemical abundances  are thus driven by rotation-induced mixing on the main sequence and convective dilution during the first dredge-up episode on the subgiant branch and early-RGB.
The predictions of our rotating models up to that phase have been extensively tested in previous papers. They account nicely for the behaviour of lithium and beryllium at the surface of Population I main-sequence and subgiant stars \citep[see \S~4.1;][]{TalCha98, TaCC10, ChaTal99, Palacios03, Pasquini04, Smiljanic09, CCNL10}. 

Rotation-induced mixing has also an impact on the internal abundance profiles of heavier chemicals  involved in hydrogen-burning at higher temperatures than the fragile Li and Be. This can be seen at the moment of the turnoff in Fig.~\ref{fig:profil_1.25_diffrot_Tof}  
for the 1.25~M$_{\odot}$ star computed for different initial rotation velocities. 

In the rotating models, the abundance gradients are smoothed out compared to the standard case: $^3$He, $^{13}$C, $^{14}$N,  and $^{17}$O diffuse outwards, while $^{12}$C and $^{18}$O diffuse inwards. However, rotation-induced mixing is not efficient enough to noticeably change the surface abundances of these elements while on the main sequence for the 1.25~M$_{\odot}$ model\footnote{The only exception is $^7$Li which is strongly depleted in the rotating case.}, 
although it sets the scene for abundance variations in latter evolution phases. 

In particular the surface abundance variations during the first dredge-up are slightly strenghtened when rotation-induced mixing is accounted for, as shown in 
Fig.~\ref{fig:surfaceabundances1p25}.  For example, more $^3$He is brought into the stellar envelope, and the post dredge-up $^{12}$C/$^{13}$C and  $^{16}$O/$^{17}$O ratios are lower than in the non-rotating case. We note, however, that neither $^{16}$O nor $^{23}$Na are affected, and that no $^{23}$Na enhancement is expected at the surface of such a low-mass star during the first dredge-up. 

\subsubsection{Red giant branch}

\begin{figure}
	\centering
		\includegraphics[angle=0,width=9cm]{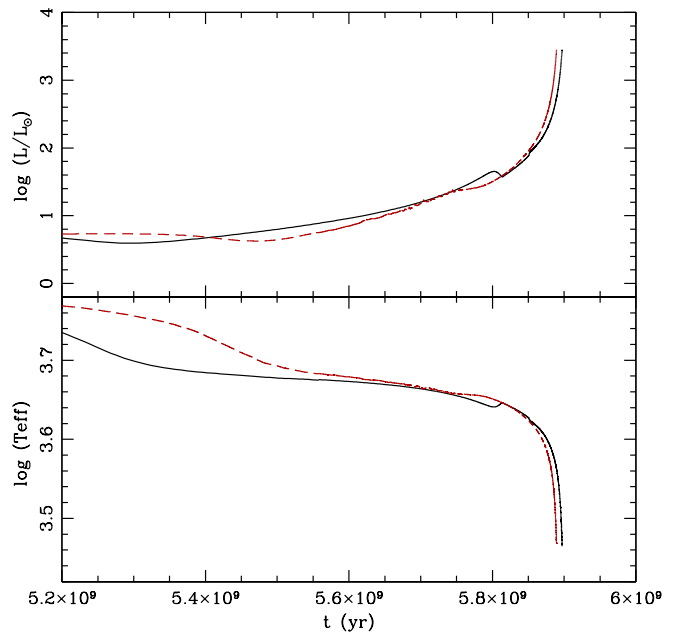}
			  \caption{Evolution with time along the RGB of the effective temperature and of the luminosity of the 1.25~M$_{\odot}$ models computed with thermohaline mixing only (black solid line) and with both thermohaline and rotation-induced mixing (initial rotation velocity of 110~km.s$^{-1}$}, red dotted line).
	\label{fig:logL_logTeff_vs_t_1p25_v0_v110}
\end{figure}

The modifications of the internal abundances due to rotation-induced mixing that we just discussed for the 1.25~M$_{\odot}$ model do induce slight variations of the overal stellar structure and of the evolutionary track. In the case of such a low-mass star, the impact on the effective temperature and luminosity along the RGB is relatively modest as can be seen in Fig.~\ref{fig:logL_logTeff_vs_t_1p25_v0_v110}.
We note that in the case of the computation without rotation (black full line) the drop in luminosity at the RGB bump (that is associated with a slight increase in Teff at $\sim 5.8 \times 10^9$ years) is clearly noticeable. In the rotating case the inflexion in luminosity at the bump is less pronounced and occurs at slightly lower luminosity. Consequently, the stellar luminosity at which the thermohaline instability reaches the convective envelope is also slightly lower in the rotating case (see Table 1). 

\begin{figure}
	\centering
		\includegraphics[angle=0,width=9cm]{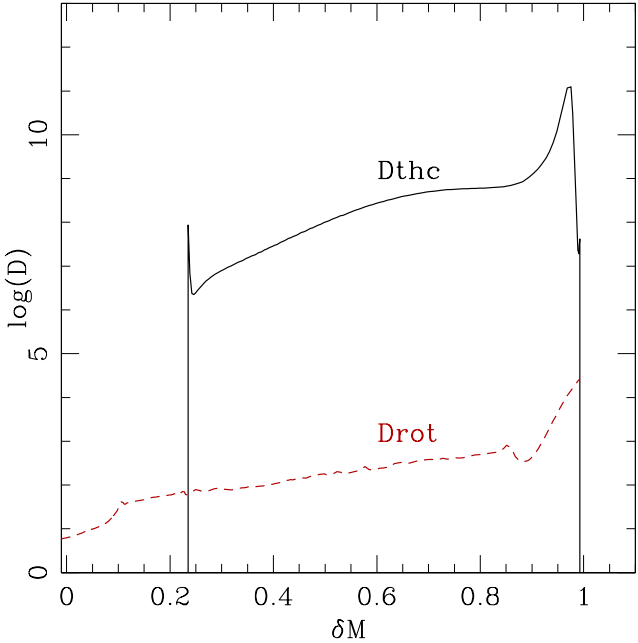}
			  \caption{Thermohaline diffusion coefficient D$_{\rm thc}$ (black solid line) and 
total rotation coefficient D$_{\rm rot}$ (red dashed line) as a function of $\delta$M for a 1.25~M$_{\odot}$ star (initial rotation velocity of 110 km s$^{-1}$) on the RGB at a  total stellar luminosity  of 105L$_{\odot}$ (i.e., at the evolution point C$_{1.25}$).
 In the current model at this precise evolution point the size of the region between $\delta$M=0 and 1 is 0.687696~R$_{\odot}$ while the  thermohaline instability extends over 0.66136~R$_{\odot}$.}
	\label{fig:Dtot_Dthc_1p25_105_deltaM}
\end{figure}

\begin{figure}
	\centering
		\includegraphics[angle=0,width=9cm]{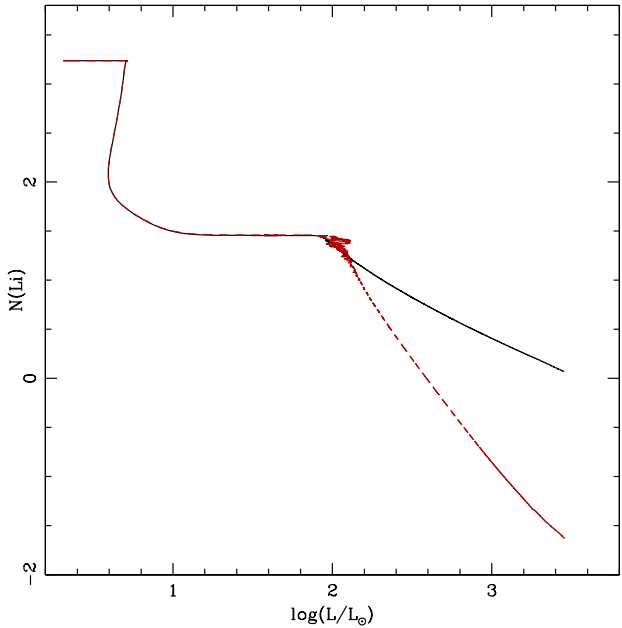}
	  \caption{Evolution of the surface abundance of $^7$Li in the 1.25~M$_{\odot}$ star up to the RGB tip when  considering thermohaline transport but no rotation-induced mixing. The black solid and red dashed curves correspond to computations performed with C$_{\rm t}=10^3$ and $10^4$ respectively.}
	\label{fig:Li_1p25_V0_Ct1e3_Ct1e4}
\end{figure}

Figure~\ref{fig:Dtot_Dthc_1p25_105_deltaM} shows, at the evolution point C$_{1.25}$ for the 1.25~M$_{\odot}$ model computed with an initial rotation velocity of 110 km.sec$^{-1}$, the diffusion coefficient associated to the  thermohaline instability, D$_{\rm thc}$ (Eq. 1 for C$_{\rm t}=1000$), and the total diffusion coefficient associated to rotation, D$_{\rm rot}$, that characterizes the transport of chemicals through the interaction of meridional circulation and shear turbulence 
\citet[][see e.g. Eqs.~5, 7 and 8 in \citealt{Palacios06}]{Zahn92}.
D$_{\rm thc}$ is five to six orders of magnitude higher than D$_{\rm rot}$, and this result is independent of the initial rotation velocity on the ZAMS. 

This confirms the finding by \citet{CaLa08,CantielloLanger2010} that in the relevant layers thermohaline mixing has much higher diffusion coefficients than rotational and magnetic instabilities\footnote{This conclusion on the magnetic diffusivity was obtained in the case of magnetic fields that are created in differentially rotating star and is not valid for magnetic stars that possess anomalous fossil fields such as the Ap star descendants discussed in \citet{ChaZah07b}.} . 
It is also perfectly consistent with the results of 
\citet[][see also \citealt{Chaname05}]{Palacios06} who studied the impact of rotation-induced mixing on the RGB for low-mass stars and showed that it cannot (with the present assumptions) account for the abundance anomalies observed in bright giants. 
These authors noted that assuming differential rotation (i.e., uniform specific angular momentum) instead of solid body rotation (i.e., uniform angular velocity) in the convective envelope along the RGB \citep[see, i.e.,][]{BrunPalacios2009} does lead to higher efficiency of the rotation-induced processes below the convective envelope. However, even in that case, Palacios and collaborators showed that the total transport coefficient associated to rotation does not rise above $10^5$ cm$^2$ s$^{-1}$ in the outer HBS, which is still much lower than the thermohaline diffusion coefficient. 

Thermohaline mixing thus governs the surface abundance variations on the upper half of the RGB as already discussed by CZ07. The corresponding predictions for the 1.25~M$_{\odot}$ model computed with an initial rotation velocity of 110~km.s$^{-1}$ can be seen in Fig.~\ref{fig:surfaceabundances1p25} (right panels). 

\subsubsection{AGB}
The 1.25~M$_{\odot}$ model with an initial rotation velocity of 110 km s$^{-1}$ was computed until the end of the superwind phase (total stellar mass and mass of the envelope being respectively equal to 0.566~M$_{\odot}$ and 0.014~M$_{\odot}$), and has undergone four thermal pulses. During the TP-AGB the behaviour of the surface abundances is similar to that discussed in \S~3.1.2.
In the present case, the maximum N(Li) value reached at the end of the TP-AGB is $\sim$ 0.8 (instead of 0.9 in the 1.25~M$_{\odot}$ non-rotating model discussed in \S~3.1; see also Table~\ref{tableNLiTPAGB}). 
Again, predictions for carbon at that phase must be taken with caution since the impact of parametric convectively induced extra-mixing is not taken into account.

\subsection{Uncertainties on the thermohaline diffusion coefficient}

\subsubsection{Size and shape of the thermohaline cells}

CZ07  performed computations for various values of the coefficient ${\rm C_ t}$ and discussed the uncertainties on the efficiency of the thermohaline mixing that are basically related to the size and shape of the thermohaline cells. Their prefered value  for the aspect ratio $\alpha$=5 (see \S~2.2) also used in the present computations 
corresponds to ``fingers" rather than ``blobs"  whose shorter mixing length would translate into smaller value (by a factor of $\sim$ 50) for the coefficient C$_{\rm t}$. Crude as it may be, this choice first advocated by \citet{Ulrich72} is supported by laboratory experiments where the instability takes the form of salt fingers  \citep{Krish03}. Also and contrary to the ``blob assumption", the ``finger prescription" leads to a very good description of the surface abundances in low-metallicity stars as shown in CZ07, as well as in solar-metallicity stars as will be discussed in \S~4.
Unfortunatly and as mentionned by \citet{Eggleton07} the 3D simulation by \citet{Eggleton06} did not have the resolution to give clues on the aspect ratio  of the fingers.

As a test we have however run a 1.25~M$_{\odot}$ model without rotation-induced mixing but with C$_{\rm t} = 10^4$ instead of the value of $10^3$ used in all the other computations presented in the present paper. The predictions for the evolution of the surface abundances up to the RGB tip are shown in Fig.\ref{fig:surfaceabundances1p25} (red lines in left panels) and \ref{fig:Li_1p25_V0_Ct1e3_Ct1e4}. As expected, increasing the thermohaline diffusion coefficient by a factor of 10 leads to faster and substantially stronger processing of material on the RGB: At the RGB tip, the surface abundances of $^3$He and of $^7$Li are reduced respectively by a factor of 2 and by $\sim$ 1.5~dex compared to the C$_{\rm t} = 10^3$ assumption. The impact on the carbon and oxygen isotopic ratios and on the surface abundance of $^{14}$N is more moderate.

Even in that case there remains enough $^3$He to drive thermohaline mixing when the star is on the TP-AGB. The Li production during that phase is higher than in the C$_{\rm t} = 10^3$ case, with the final surface abundance N(Li)=2 instead of 0.9. The final $^3$He abundance and carbon isotopic ratio are $1.8 \times 10^{-4}$ (in mass fraction) and 8 respectively (instead of  $3.15 \times 10^{-4}$ and 9.65).

\subsubsection{Atomic diffusion}

In the present computations we have not included the effects of atomic diffusion. 
In particular we do not consider radiative levitation that may lead to accumulation of heavy elements and thus to thermohaline instability in the outer layers of peculiar, slowly-rotating main sequence A-type stars \citep [see] []{Theado09}.  This simplification has no effect on our conclusions, since the thermohaline instability induced by iron accumulation affects only the very external regions of these atypical main sequence stars and has no direct impact on the nuclear burning occuring much deeper inside the star, nor on the RGB chemical properties.

We do not consider either the effect of atomic diffusion on the RGB, although \citet{Michaud10} pointed out that at that phase $^4$He gravitational settling may eventually lead to a larger $\mu$-inversion than $^3$He-burning on the outskirts of the HBS. In their computations, however, thermohaline mixing is not taken into account. Consequently the effects of concentration variations on $\mu$ they get  from pure atomic diffusion are maximum compared to reality where thermohaline mixing (induced by $^3$He-burning and eventually by $^4$He-settling) counteracts atomic diffusion. Michaud and collaborators have estimated that a value of D$_{\rm thc}$ of the order of 10$^7$ cm$^2$ s$^{-1}$ is able to substantially reduce (by a factor of 10) the small gradients of He caused by atomic diffusion on the RGB for a 0.95M$_{\odot}$, Z=0.004 model. Given that this number is smaller than D$_{\rm thc}$ obtained in our RGB models (see Fig.~\ref{fig:Dthc_X_1p25} and \ref{fig:Dtot_Dthc_1p25_105_deltaM}), we can safely assume that the effects of atomic diffusion must be wiped out by turbulence and that 
 $^3$He-burning is the dominating process inducing  thermohaline instability between the HBS and the convective envelope in RGB stars. We are aware that some $^4$He settling may remain even under the counteracting action of thermohaline mixing, although this should be confirmed by computations that are out of the scope of the present paper. One may note, however, that this should simply slightly re-inforce the $\mu$-inversion induced by $^3$He-burning (although to a much lower extent than in Michaud's computations), and thus strengthen the thermohaline transport compared to the present models.

\subsection{Low-mass stars more massive than $\sim$1.5~M$_{\odot}$}

Table~1 gives the luminosity of the bump as well as the luminosity L$_{\rm c, th}$ at which the thermohaline instability ``contacts" the base of the convective envelope for all the low-mass stellar models computed both without and with rotation for the present study. As we have just seen, for RGB stars less massive than $\sim$1.5~M$_{\odot}$, thermohaline mixing starts changing the surface abundances soon after the bump. 
However, for RGB stars with initial mass higher than 1.5~M$_{\odot}$ computed without rotation-induced mixing, the thermohaline instability is long quenched into a very thin region, and is able to connect the external wing of the HBS with the convective envelope only when the star reaches already very high luminosity, close from the RGB tip. This is consistent with the finding by \citet{CaLa08,CantielloLanger2010} of an upper mass limit for efficient thermohaline mixing in non-rotating low-mass RGB stars. However, as we shall see below, this conclusion does not hold anymore when one considers the impact of stellar rotation.

\begin{figure}
	\centering
		\includegraphics[angle=0,width=9cm]{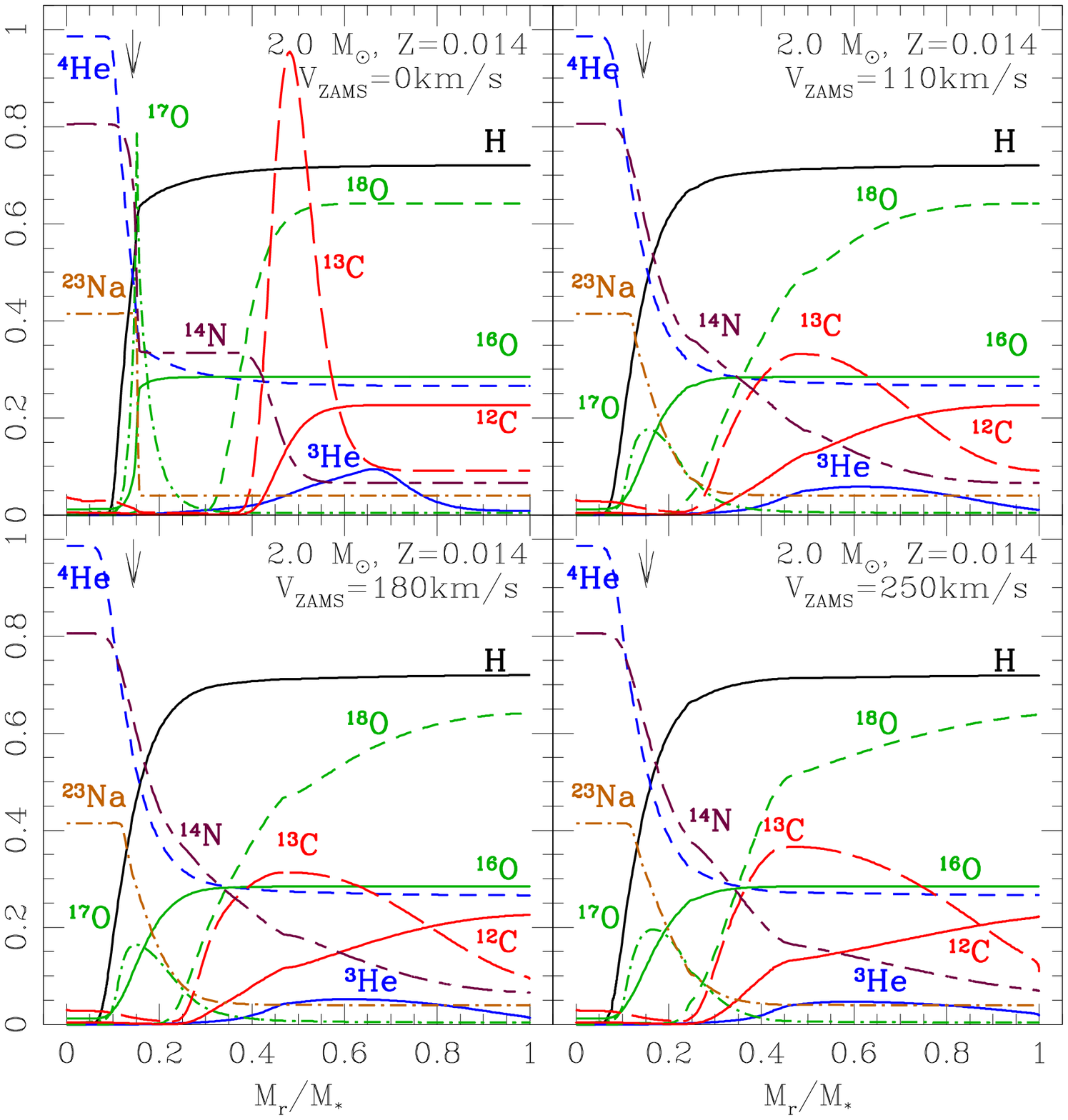}
	  \caption{Same as Fig.~\ref{fig:profil_1.25_diffrot_Tof} for the 2.0~M$_{\odot}$ star computed for different initial rotation velocities, as indicated. The mass fractions are multiplied by 100 for $^3$He, $^{12}$C, and $^{14}$N, by 2500 for $^{13}$C, by 50, 1100, and 5$\times 10^4$ for $^{16}$O, $^{17}$O, and $^{18}$O respectively, and by 1500 for $^{23}$Na.  The vertical arrows show, in all cases, the maximum depth reached by the convective envelope at its maximum extent during the first dredge-up.}
	\label{fig:profils_abon_rotdiff_2}
\end{figure}

 \begin{figure}
	\centering
		\includegraphics[angle=0,width=9cm]{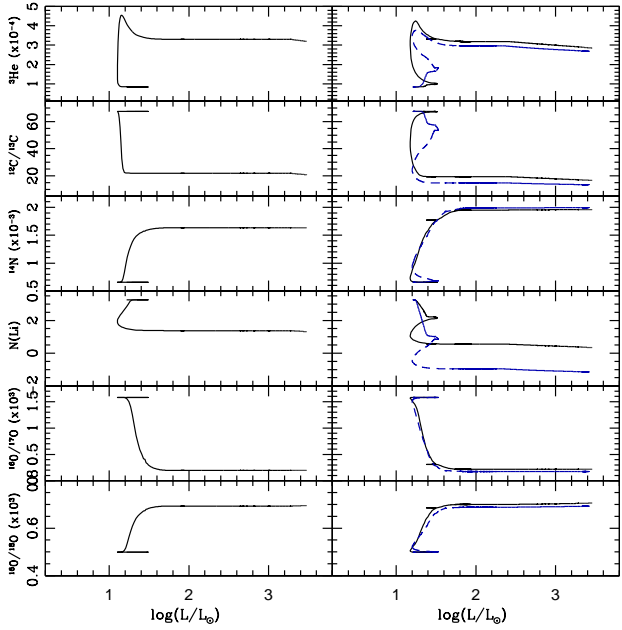}
	  \caption{Evolution of the surface abundances of $^3$He, $^7$Li, $^{14}$N, and of the carbon and oxygen isotopic ratios for the 2.0~M$_{\odot}$ star from the zero age main sequence up to the RGB tip. {\it(Left)} Model including thermohaline mixing only. {\it (Right)} Models including thermohaline and rotation-induced mixing for different initial rotation velocities (110 and 250 km s$^{-1}$, solid black and dashed blue lines respectively). In all cases C$_{\rm t}=10^3$.}
	\label{fig:surfaceabundances2p0}
\end{figure}

Let us now discuss indeed the case of a $2.0$~M$_{\odot}$ star which interior chemical structure at the turnoff is shown in  Fig.\ref{fig:profils_abon_rotdiff_2}  for different initial velocities. 
 As for the 1.25~M$_{\odot}$ models discussed before, rotation-induced mixing  smoothes the abundance profiles inside the star and in the present case it already leads to variations of the surface abundances on the main sequence that are stronger for higher initial rotation velocities (this can be seen by looking at the abundance values at M$_r$/M$_*$=1 in the various panels of Fig.\ref{fig:profils_abon_rotdiff_2};  \cite[see also e.g.][]{MeynetMaeder02}).  
 Note also that the maximum value of $^3$He at the peak is much lower than in the 1.25~M$_{\odot}$, because on the main sequence the $2.0$~M$_{\odot}$ burns hydrogen mainly through the CNO cycle rather than through the pp-chains. The more massive star thus dredges less $^3$He on the subgiant branch as can be seen by comparing Fig.~\ref{fig:surfaceabundances1p25} and \ref{fig:surfaceabundances2p0}.

Figure\ref{fig:surfaceabundances2p0} shows the evolution of the surface abundances of $^3$He, $^7$Li, $^{14}$N, and of the $^{12}$C/$^{13}$C,  $^{16}$O/$^{17}$O, and $^{16}$O/$^{18}$O ratios as a function of luminosity up to the RGB tip in 2.0~M$_{\odot}$ models computed without or with rotation (left and right panels respectively). At this stellar mass some $^{23}$Na produced during the main sequence is dredged-up (not shown in Fig\ref{fig:surfaceabundances2p0} but see Table~2). 
All these quantities at the end of the first dredge-up are slightly affected by rotation-induced mixing that changed the abundance profiles while the star was on the main sequence (e.g., the post dredge-up value of  the carbon isotopic ratio is lower when faster rotation is accounted for). 
Furthermore, we note that the changes in surface abundances due to thermohaline mixing start at much lower luminosity (closer from the bump) on the RGB than in the non-rotating case (see also Table~1). Overall, the total abundance variations at the tip of the RGB are stronger for higher initial rotation velocity.

Thus and contrary to the conclusion by \citet{Cantiello07} and \citet{CantielloLanger2010}, we find no mass limit (in the case of low-mass stars that ignite He in a degenerate core) for thermohaline mixing  to change the surface abundances on the RGB. However, the global efficiency of this process increases when one considers less massive stars at a given metallicity, or more metal-poor stars at a given stellar mass. As discussed previously, this results from the combination of several factors like the thermohaline diffusion timescale compared to the secular timescale, the compactness of the HBS and of the thermohaline unstable region, and the amount of $^3$He available to power the thermohaline instability.

In this mass range thermohaline mixing also leads to Li production during the TP-AGB (see Table~\ref{tableNLiTPAGB}). 
In the 2.0~M$_{\odot}$ model computed with an initial rotation velocity of 110 km s$^{-1}$, N(Li) increases from a value of 0.1 on the early AGB up to $\sim$ 1.5 at the end of the TP-AGB (the envelope mass is then $\sim$ 0.14~M$_{\odot}$, and the star has undergone 11 thermal pulses). Meanwhile the $^3$He surface abundance has slightly decreased from $2.74 \times 10^{-4}$ to $2.67 \times 10^{-4}$.

We wish to emphasize an interesting result obtained for the 2.0~M$_{\odot}$ model that was computed up to the AGB tip with thermohaline mixing but without rotation. This model did undergo 11 thermal pulses in total. 
After the 9th thermal pulse third dredge-up occurred, that slightly increased the $^{12}$C surface abundance as well as the carbon isotopic ratio. During the following interpulse phase this ratio was slightly lowered under the effect of thermohaline mixing. Quantitatively, the carbon isotopic ratio increased from 19.6 to 21.33 between the end of the second dredge-up  and the AGB tip (see also Fig.\ref{fig_c1213_PNe} and discussion in \S~4.2). This could indicate that thermohaline mixing does favour the occurrence of third dredge-up. This will be investigated further in a future paper.

 \subsection{Theoretical predictions for intermediate-mass stars}
 
\begin{figure}
	\centering
		\includegraphics[angle=0,width=9cm]{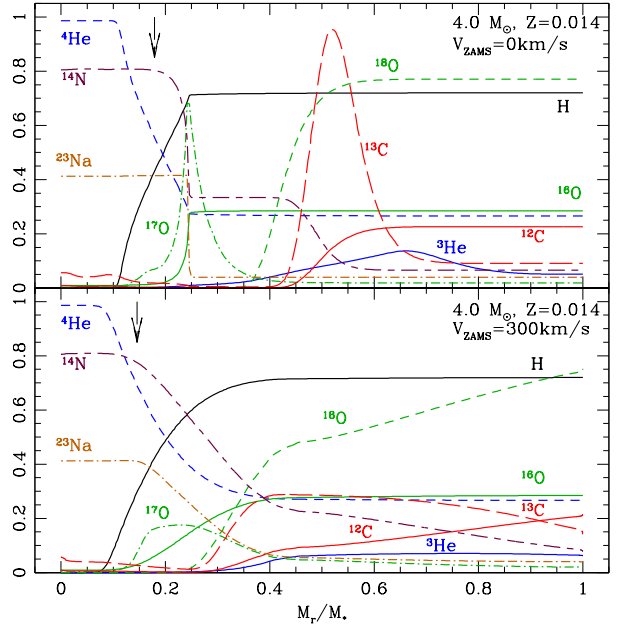}
	  \caption{Chemical structure at the turnoff of the 4.0~M$_{\odot}$ star computed without rotation and with rotation for an initial rotation velocity of 300 km s$^{-1}$ (top and bottom panels respectively).
	  The mass fractions are multiplied by 600 for $^3$He,  by 100 for $^{12}$C and $^{14}$N, by 2500 for $^{13}$C, by 50, 5000, and  6$\times 10^4$ for $^{16}$O, $^{17}$O, and $^{18}$O respectively, and by 1500 for $^{23}$Na. 
	   The vertical arrows show, in both cases, the maximum depth reached by the convective envelope at its maximum extent during the first dredge-up.}
	\label{fig:profils_abon_rotdiff_4}
\end{figure}

By definition, intermediate-mass stars are objects that ignite central helium-burning in a non-degenerate core at relatively low luminosity on the RGB, well before the HBS reaches the mean molecular weight discontinuity caused by the first dredge-up.  In other words, these objects do not go through a bump on their short ascend of the RGB, and thus do not undergo thermohaline mixing at that phase. 

We should note, however, that as in the previous cases, rotation-induced mixing can not be neglected from the whole picture. We refer to \citet{Eggenbergeretal10} for a discussion of the global effects of rotation on the evolution and asterosismic properties of intermediate-mass red giants \cite[see also e.g.][]{MeynetMaeder02}. 
Its impact on the chemical structure of a 4.0~M$_{\odot}$ star at turnoff can be seen in Fig.~\ref{fig:profils_abon_rotdiff_4}. Note that in this mass range the base of the convective envelope reaches the $^{23}$Na plateau during the first dredge-up, leading to an increase of the surface abundance of this element both in the non-rotating and rotating cases. 

Overall rotation-induced mixing leads to stronger modifications of the stellar chemical properties when the star becomes a giant as its convective envelope dredges-up nuclearly processed material. At the end of the dredge-up for the 4~M$_{\odot}$ models, the surface $^3$He abundance is $1.1 \times 10^{-4}$ and $8.4 \times 10^{-5}$ in the non-rotating and rotating (initial rotation velocity of 300 km s$^{-1}$) models respectively, while the carbon isotopic ratio is 20.5 or 14 respectively, and N(Li) is 1.3 or -1.8.

We computed the first 11 thermal pulses for the 4~M$_{\odot}$ models without and with rotation-induced mixing, including thermohaline mixing in both cases. A strong Li increase at the stellar surface is obtained during the first thermal pulses, and then lithium production levels off at a value of the order of N(Li)=2.2. 
This agrees with \citet{Stancliffe10} predictions.  
Again, a detailed exploration of the TP-AGB phase for intermediate-mass stars is postponed to a further paper.

%=========================================================== Partie 5 Comparaison avec Obs =========================================================

\section{Comparison with observations}

\begin{figure}
	\centering
		\includegraphics[angle=0,width=9cm]{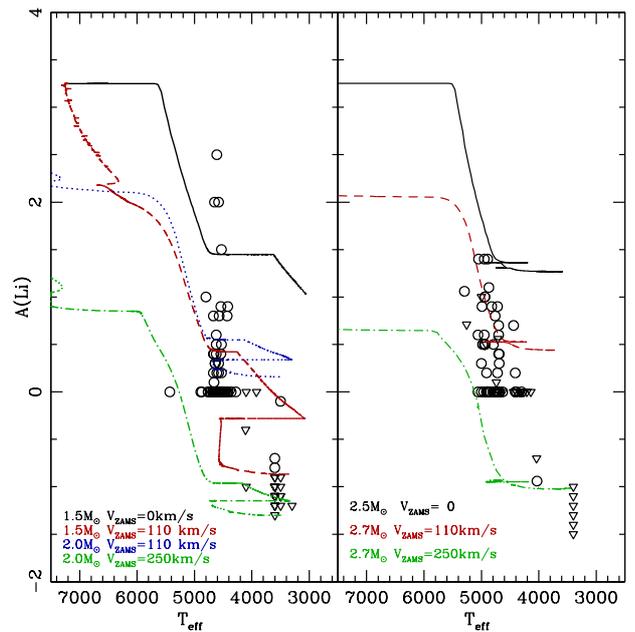}
	  \caption{Lithium data for field evolved stars from the sample by Charbonnel et al. (in prep., see the text) that are segregated according to their mass (left and right panels include respectively sample stars with masses lower and higher than 2~M$_{\odot}$; Li detections and upper limits are shown as circles and triangles respectively). 
	  Theoretical lithium evolution is shown from the ZAMS up to the end of the early-AGB. 
	  Various lines correspond to predictions for stellar models of different masses computed without or with rotation as indicated, and with thermohaline mixing in all cases (with C$_{\rm t}=10^3$).
	  }
	\label{fig:Lagardeetal_posterIAU268_figb}
\end{figure}

We now test the theoretical predictions of our models with respect to observations of relevant chemical elements in stars at different evolution stages. 
In Table \ref{tablesurfabundances} we give the surface carbon isotopic ratio as well as surface abundances of $^{7}$Li, $^{9}$Be, [C/Fe], [N/Fe], and [Na/Fe], at the end of the first dredge-up, at the RGB tip, and at the end of the second dredge-up, 
for each of the models we have computed. In this table and in the following figures all the predictions correspond to models computed with a value of C$_{\rm t} = 10^3$ for thermohaline mixing (without or with rotation-induced mixing). As underlined in \S~3, we did not include parametric convectively induced extra-mixing during the TP-AGB phase so that our models are not expected to undergo third dredge-up and to mimic carbon-rich stars. 

\subsection{Lithium behaviour}

\subsubsection{Lithium destruction}

As already mentioned the predictions of the present rotating models 
have been successfully compared to Li and Be observations along the whole evolutionary sequence of the Galactic open cluster IC~4651 (turnoff mass 1.8~M$_{\odot}$) by \citet [see their Fig.10 to 14] {Smiljanic10}. They account very nicely for all the Li and Be features observed in this cluster.

Here we use as additional constraints
Li observations that we performed for a large sample of field red giant stars (subgiant, RGB, and early-AGB stars) with metallicities around solar. All sample stars have Hipparcos parallaxes so that their mass and evolutionary status could be relatively well determined (Charbonnel et al. in preparation). In Fig.\ref{fig:Lagardeetal_posterIAU268_figb}  they are distinguished with respect to their mass (less or  more massive than 2~M$_{\odot}$ in the left and right panels respectively). 

Let us consider first the stars with initial masses lower than 2~M$_{\odot}$, whose Li properties  are compared with predictions for the 1.5 and 2~M$_{\odot}$ models (left panel of Fig.\ref{fig:Lagardeetal_posterIAU268_figb}). 
The theoretical Li behaviour is relatively straightforward. 
On the main sequence and on the early-RGB, rotation-induced mixing leads to stronger Li depletion than in the standard case (compare e.g. the red curve with the black one); 

in this mass range indeed standard models predict no Li depletion on the main sequence and a N(Li) of the order of 1.5 at the end of the first dredge-up, which is at odds with the data. After the end of the first dredge-up (Teff $\sim$ 4800~K), the theoretical Li abundance remains temporarily constant as the convective envelope withdraws in mass. When thermohaline mixing becomes efficient (Teff $\sim$ 4200~K), the theoretical Li abundance drops again in drastic manner (while it would stay constant in the standard case). 
After the star has reached the RGB tip its effective temperature increases (up to $\sim$ 4800~K) as it settles on the clump, before decreasing again when the star starts climbing the early-AGB. The second dredge-up that occurs then leads to a final decrease of N(Li). On this graph we do not plot the Li increase that is predicted to occur during the TP-AGB phase at a Teff of $\sim$ 3200~K due to thermohaline mixing, and which is discussed in \S~4.1.2.  
As can be seen in Fig.\ref{fig:Lagardeetal_posterIAU268_figb}, the present predictions are in perfect agreement with the data all along the evolutionary sequence and explain very well the upper limits observed for the brightest sample giant stars. The observed Li dispersion at a given effective temperature reflects dispersion in the initial rotation velocity and in the initial stellar mass \citep[see also] [] {ChaTal99, Palacios03, Smiljanic10}. 

The case of the more massive stars, whose Li observational behaviour is compared to predictions for the 2.5 and 2.7~M$_{\odot}$ models (right panel of Fig.\ref{fig:Lagardeetal_posterIAU268_figb}) is even more simple. In these objects indeed no thermohaline mixing occurs on the too short RGB, and rotation-induced mixing alone explains very well the data. 

\subsubsection{Lithium production}

\begin{figure}
	\centering
		\includegraphics[angle=0,width=9cm]{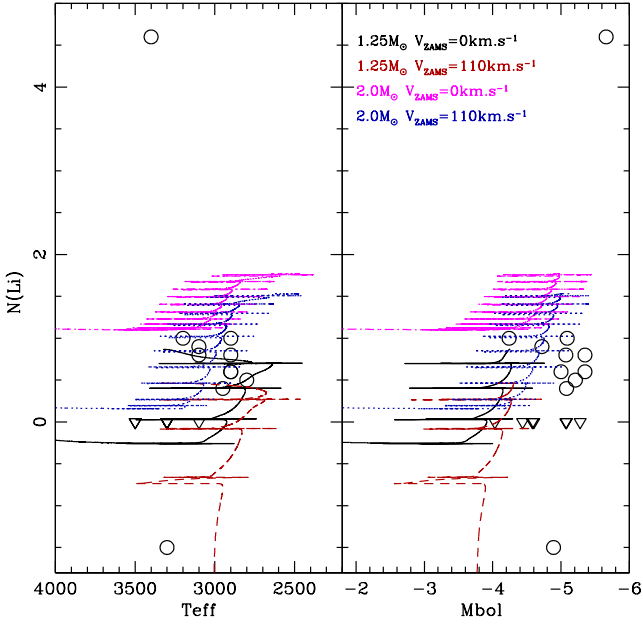}
	  \caption{Lithium observations in \citet{UttenthalerLebzelter10} sample of oxygen-rich variables belonging to the Galactic disk (circles and triangles are for abundance determinations and upper limits respectively) as a function of effective temperature and bolometric magnitude. 
	  Typical error bars are indicated.
	   The star with the highest Li abundance is W441 Cyg (see text).
	  Theoretical lithium evolution is shown from the early-AGB up to the end of the TP-AGB. 
	  Various lines correspond to predictions for stellar models of different masses computed without or with rotation as indicated, and with thermohaline mixing in all cases (with C$_{\rm t}=10^3$).
	  }
	\label{fig:TP_Lithium_Uttenthaler}
\end{figure}

In all the models that we have computed along the TP-AGB, non negligible fresh lithium production is obtained, although the total stellar yields remain negative. Table~\ref{tableNLiTPAGB} gives the final N(Li) value and the total Li yield for those models.
 In all cases thermohaline transport is responsible for this strong Li enrichment.

	\begin{table}
\caption{Surface lithium abundance after the second dredge-up and at the end of the TP-AGB phase, and total lithium yield.}          
  \label{tableNLiTPAGB}    
   \centering                                                             
    \begin{tabular}{c | c | c | c | c | c }           
     \hline                         M & & $V_{zams}$& N(Li) & N(Li) & Yield \\
   (M$_{\odot})$ & & $(km.s^{-1})$ & 2DUP & tip AGB & (M$_{\odot}$) \\
   \hline      1.0 & th       & 0     & -1.86  & -0.3 & -5.17 $\times 10^{-9}$ \\
   \hline
   1.1 & th       & 0     & -0.97  & 0.25 & -5.81 $\times 10^{-9}$ \\
   \hline
   1.25 & th       & 0     & -0.25  & 0.87 & -6.75 $\times 10^{-9}$ \\
            & th+rot & 80 & -3.93 & 0.46 & -6.79 $\times 10^{-9}$  \\
           & th+rot & 110 & -3.91 & 0.85 & -6.79 $\times 10^{-9}$  \\
   \hline
   1.5 & st          & 0      &  1.38     & 1.38 &    -8.25 $\times 10^{-9}$ \\
       & th       & 0      & 0.63      & 1.49 & -8.25 $\times 10^{-9}$ \\
            & th+rot & 110 & -0.8       & 1.04 & -8.35 $\times 10^{-9}$  \\    \hline
   1.9 & th           & 0    & 1.05 & 1.8 &-1.0 $\times 10^{-8}$  \\
   \hline
   2.0 &  th           & 0    & 1.1   & 1.8 & -1.12 $\times 10^{-8}$ \\
           & th+rot    & 110 & 0.16  & 1.52 & -1.14 $\times 10^{-8}$ \\       \hline
       \end{tabular}
\end{table}

The evolution of the surface abundance along the TP-AGB is shown as a function of effective temperature and bolometric magnitude in Fig.\ref{fig:TP_Lithium_Uttenthaler} for the 1.25 and 2.0~M$_{\odot}$ models computed without and with rotation, and with thermohaline mixing in both cases (with C$_{\rm t}=10^3$). Li production starts at slightly higher effective temperature and luminosity for the more massive star.  
 
Predictions are compared with lithium values in the sample of  low-mass oxygen-rich AGB variables    belonging to the Galactic disk studied by  \citet{UttenthalerLebzelter10}. 
Theoretical Li production sets in above a lower luminosity limit which agrees with the observational Mbol threshold. 
Models are found to fit very nicely the lithium behaviour. However, they can not account for the very high Li abundance (3.1 to 4.6 depending on the model atmosphere) of the star V441 Cyg, which may rather be an intermediate-mass AGB star undergoing hot bottom burning \citep[see discussion in][]{UttenthalerLebzelter10}.

\subsection{Carbon isotopic ratio}
 
 \begin{table}
\caption{ References for the abundance studies in Galactic open clusters used in the comparisons with model predictions.  \citet{Brown87}: B87;  \citet{Gilroy89}: G89; \citet{GiBr91}: GB91; \citet{Hamdani00}: H00; \citet {Jacobson07, Jacobson08, Jacobson09} : J07, J08, J09, J10; \citet{Mikolaitisetal10}: M10; \citet{Smiljanic09}: S09; \citet{Yong05}: Y05. The red turnoff masses given in the second column were estimated using the WEBDA database and Geneva isochrones (see text).
Listed in the last column are the symbol colours used in Fig.~17 and Fig.~19 to 21.}
\label{openclusters}      
\centering                                                            
	\begin{tabular}{c | c | c | c | c | c }          
	\hline                      
	open     & M$_{\rm red TO}$ & $^{12}$C/$^{13}$C    & N/C & Na  & symbol  \\
	cluster & (M$_{\odot})$         &  &  & & colour  \\
	\hline      
	M67            & 1.5    & GB91 & B87 & Y05 & black \\
	NGC 752   &    2.0  & G89    & --      & -- & merlot \\
	NGC 6939 & 1.57 & --         &  --      & J07 & light green \\                       
	NGC 7142 & 1.67 & --         & --       & J08 & yellow \\
	NGC 3680 & 1.70 & --         & --       & S09 & pink \\  
	NGC 2141 & 1.73 & --         & --       & J09 & sauvignon \\
	NGC 2360 & 1.98 & S09    & S09  & S09 & cyan \\ 
	NGC 2158 & 2.04 & --         & --      & J09 & lime \\
	NGC 1883 & 2.08 & --         & --      & J09 & grey \\
	NGC 5822 & 2.14 & S09    & S09  & S09 & orange \\
	IC 4756      & 2.31 & S09    & S09  & J07, S09 & green \\  
	NGC 6134 & 2.31 & S09, M10    & S09  & S09, M10 & red \\
	NGC 2447 & 2.74 & S09    & S09  & S09 & olive \\  
	NGC 6633 & 2.74 & S09         & S09  & H00  & light blue \\
	NGC 1817 & 2.82 & --         & --       & J09 & brown \\
	IC 2714      & 2.85 & S09    & S09  & S09 & blue purple \\
	NGC 3532 & 2.96 &  S09   & S09  & S09 & blue \\  
	NGC 6281 & 3.09 & S09    & S09  & S09 & magenta \\
	\hline 
	\label{table:data_openclusters}
		\end{tabular}
\end{table}
	
\begin{figure}
	\centering
		\includegraphics[angle=0,width=9cm]{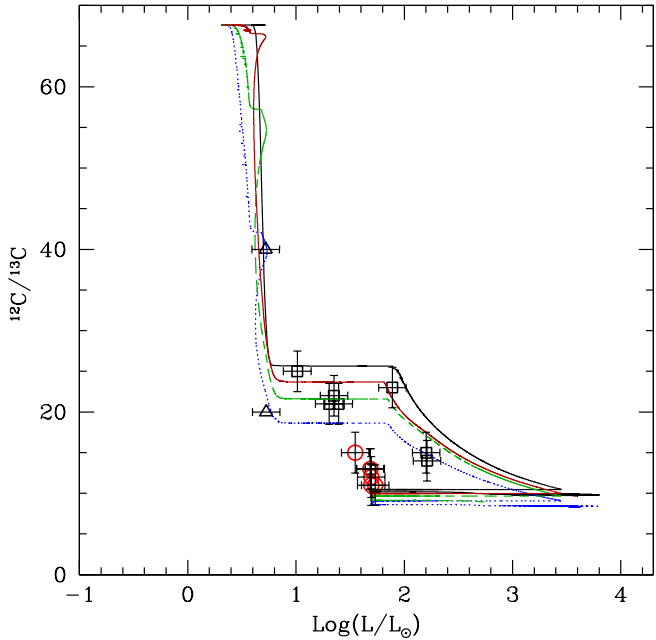}
			  \caption{Evolution of the surface $^{12}$C/$^{13}$C value as a function of stellar luminosity for the 1.25M$_{\odot}$ models including thermohaline instability and rotation-induced mixing (for initial rotation velocities of 50, 80, and 110 km.s$^{-1}$ shown as solid red, dashed green, and dotted blue lines respectively). The non rotating case is also shown (black solid line).
			    Observations  along the evolutionary sequence of the open cluster M67 are from \citet{GiBr91}. The triangle is for a subgiant star for which only a lower value could be obtained, while black squares and red circles correspond respectively to RGB and clump stars.}
	\label{fig:c1213_1.25_diffrot_gilM67}
\end{figure}

As discussed in the introduction, the behaviour of the carbon isotopic ratio is the best indicator of non-standard transport processes in evolved low-mass stars. 
This quantity has been determined in a large number of stars in Galactic open clusters. 
The references for the studies we consider here are listed in Table 4 where we also give the cluster red turnoff masses \citep[see][]{Meynetetal93} derived from the WEBDA database when using the Geneva isochrones \citep{Schalleretal1992}. 

Figure \ref{fig:c1213_1.25_diffrot_gilM67}  displays observations of the $^{12}$C/$^{13}$C ratio in subgiant, RGB, and clump stars of the open cluster M67 (turnoff mass $\sim$ 1.2~M$_{\odot}$ according to  \citet {GiBr91}, and 1.5~M$_{\odot}$ according to WEBDA, see above) by \citet {GiBr91}. The data are compared to the predictions of our 1.25M$_{\odot}$ models computed for three different initial velocities and including thermohaline mixing (the non-rotating case is also shown). We see that the theoretical and observational behaviours are in complete agreement all along the evolutionary sequence. While the dispersion for stars that have not yet reached the RGB bump (i.e., with Log(L/L)$_{\odot}$ between $\sim$ 0.7 and 1.8) reflects only the dispersion in the initial rotation velocity, 
explaining the data for more evolved stars requires the occurrence of thermohaline mixing as predicted by the models. 

\begin{figure}
	\centering
		\includegraphics[angle=0,width=9cm]{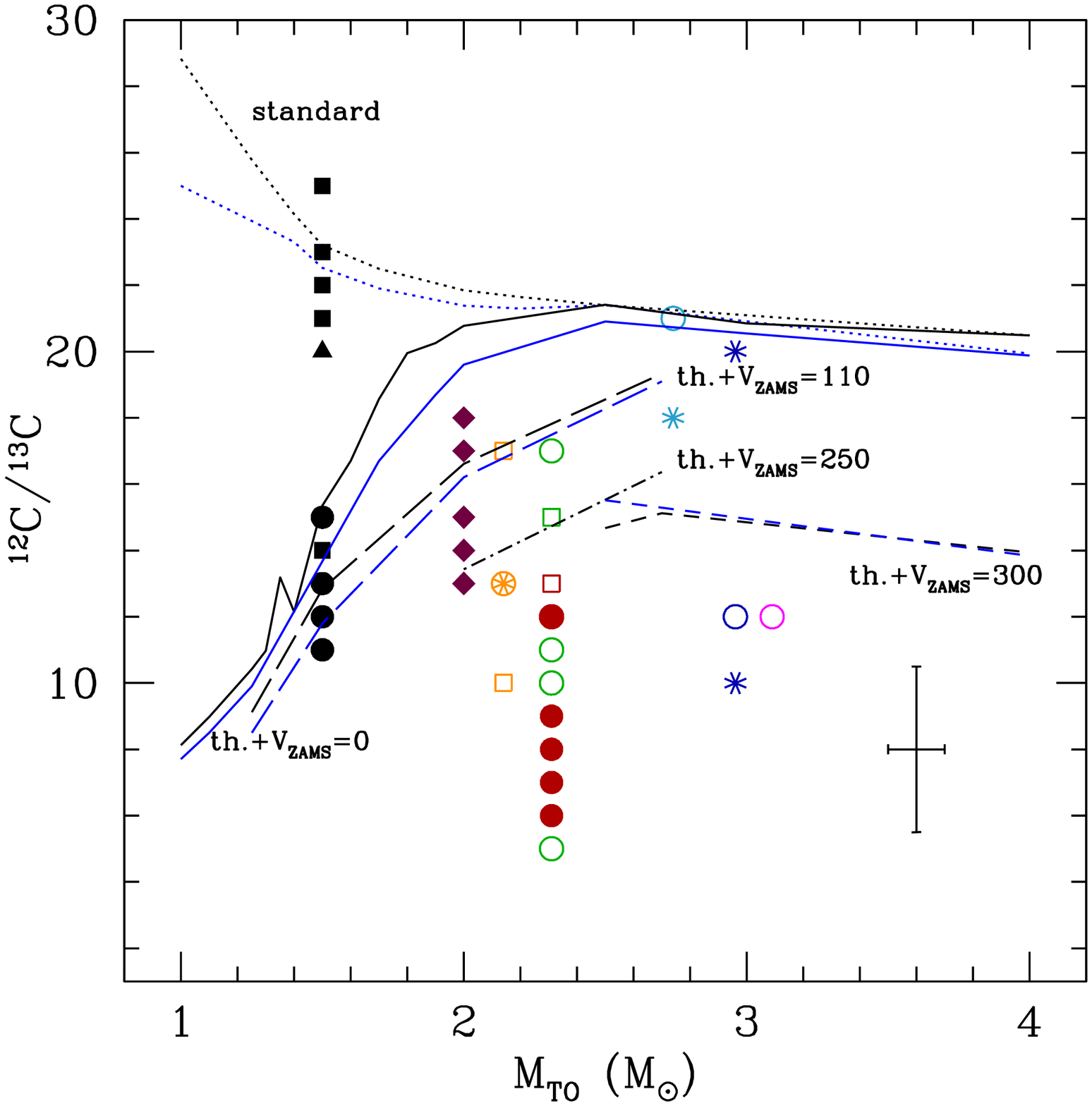}
		  \caption{Observations of $^{12}$C/$^{13}$C in evolved stars of Galactic open clusters by  \citet[open symbols]{Smiljanic09}, \citet{Gilroy89}, \citet{GiBr91}, and \citet{Mikolaitisetal10} as a function of the turnoff mass of the corresponding host cluster that can be identified thanks to the colours of the symbols (see text and Table~\ref{table:data_openclusters}).
	  Squares, circles, and asteriscs are for RGB, clump, and early-AGB stars respectively, while diamonds are for stars from \citet{Gilroy89} sample with doubtful evolutionary status; triangles are for lower limits. 
	  	  A typical error bar is indicated. 
	  Theoretical predictions are shown at the tip of the RGB and after completion of the second dredge-up (black and blue lines respectively). Standard models (no thermohaline nor rotation-induced mixing) are shown as dotted lines, models with thermohaline mixing only (V$_{\rm ZAMS}$=0) as solid lines, and models with thermohaline and rotation-induced mixing for different initial rotation velocities as indicated as long-dashed, dot-dashed, and dashed lines.}
	\label{fig:c1213_Mto}
\end{figure}

We now compare the predictions of our models over the 1.0-4.0~M$_{\odot}$ range with carbon isotopic ratios in open clusters of different turnoff masses. 
The data shown in Fig.~\ref{fig:c1213_Mto} are from \citet{GiBr91} for M67,  from \citet{Gilroy89} for her open clusters with turnoff masses below 1.7~M$_{\odot}$ (i.e., NGC~752), from the more recent study by \citet{Smiljanic09} for nine pen clusters with  turnoff masses above 1.7~M$_{\odot}$ (IC~2714, IC~4756, NGC~2360, NGC~2447, NGC~3532, NGC~5822, NGC~6134, NGC~6281, NGC~6633), and from \citet{Mikolaitisetal10} for NGC 6134.
Individual stars are attributed the red turnoff mass of their host cluster determined as described above (see Table~4).
Indications on their evolutionary status, when available,  are given in the plot ( squares, circles,  and asteriscs are for RGB, clump, early-AGB stars respectively, while diamonds are for stars with uncertain evolutionary status). 
Model predictions are shown both at the tip of the RGB and at the end of the second dredge-up\footnote{Note that with the present computations without parametric convective overshoot during the thermal pulses, the carbon isotopic ratio is only very slightly modified during the TP-AGB phase compared to its value at the end of the second dredge-up.} (black and blue lines respectively) for different assumptions. 
Dotted lines correspond to standard models computed without thermohaline mixing nor rotation; those account only for the upper envelope of the data. 
Solid lines correspond to models computed with thermohaline mixing only. They reproduce very well the $^{12}$C/$^{13}$C behaviour for stars with initial masses lower than $\sim$1.7~M$_{\odot}$. In this mass range rotation-induced mixing leads only to slightly lower values as shown by the long-dashed lines for an initial velocity of 110 km s$^{-1}$.  Note that the squares at M$_{\rm turnoff}$=1.5~M$_{\odot}$ correspond to the M67 stars that have not yet reached the RGB bump (see Fig.\ref{fig:c1213_1.25_diffrot_gilM67}), which explains why they lie between the standard and the thermohaline curves. For stars with masses between $\sim$1.7 and 2.2M$_{\odot}$ both thermohaline and rotation-induced mixing are required to fit the data.
For more massive stars thermohaline mixing plays no role but the observational uncertainties allow the data to be well accounted for when rotation-induced mixing is taken into account (dot-dashed and dashed lines for initial velocities of 250 and 300 km s$^{-1}$ respectively).   

\begin{figure}
	\centering
		\includegraphics[angle=0,width=9cm]{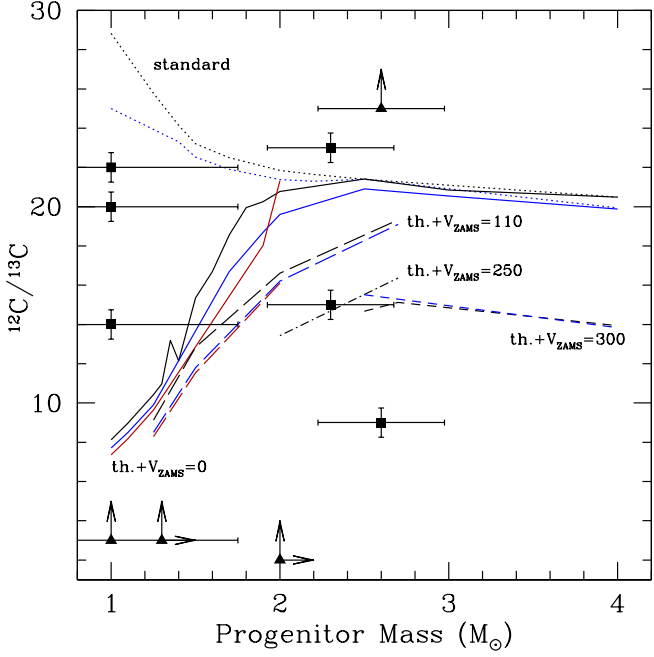}
	 \caption{ Observations of $^{12}$C/$^{13}$C in planetary nebulae as a function of the progenitor mass \citep{Pallaetal00}. Squares and triangles are respectively  for actual determinations and  lower limits of the carbon isotopic ratio.  Black and blue theoretical lines have the same meaning as in Fig.\ref{fig:c1213_Mto}, while  red curves show the model predictions at the AGB tip for the low-mass models that were computed up to that phase (solid and dashed red curves are respectively for the models with thermohaline mixing only  and for the models with thermohaline and rotation-induced mixing). }
	 \label{fig_c1213_PNe}
\end{figure}

Finally we compare in Fig.\ref{fig_c1213_PNe} our predictions for the carbon isotopic ratio to its determination in a sample of planetary nebulae obtained by means of millimeter wave observations of $^{12}$CO and $^{13}$CO \citep{Pallaetal00}. 
The abcissa is the progenitor mass derived by Palla and collaborators; this quantity is highly uncertain as shown by the error bars. 
Since the formation of the planetary nebula occurs at the AGB tip, the data should be compared to the model predictions at the end of the superwind phase as plotted in red. 
As explained in \S~3 only some of our low-mass models (with initial mass $\leq$ 2~M$_{\odot}$) were computed up to that phase. For these objects thermohaline mixing was found to slightly lower the carbon isotopic ratio during the thermal pulse phase, except in the 2~M$_{\odot}$ model with thermohaline mixing and no rotation that has undergone third dredge-up from the 9th pulse on (see \S~3.4); in that case the carbon isotopic ratio at the AGB tip is slightly higher than at the end of the second dredge-up.
Overall the comparison between the models and the data turns out to be quite satisfactory. 
Two of the planetary nebulae with low-mass progenitors (namely NGC~6781 and M~1-17) actually exhibit relatively high carbon isotopic ratio. Although the uncertainty on the initial stellar mass of these objects allows the data to be well accounted for by the thermohaline models (both with and without rotation), it could also be that in this couple of stars thermohaline mixing was inhibited by strong fossil magnetic fields as suggested by \citet{ChaZah07b}. In this context it would be extremely valuable to look for magnetic fields in these two possible ``thermohaline deviant stars".
On the other hand, computations are now needed to estimate the combined effect of third dredge-up, hot bottom burning, and thermohaline mixing during the TP-AGB phase for stars more massive than 2~M$_{\odot}$. This is work is in progress.

\subsection{Nitrogen, sodium, and oxygen isotopes}

\begin{figure}
	\centering
		\includegraphics[angle=0,width=9cm]{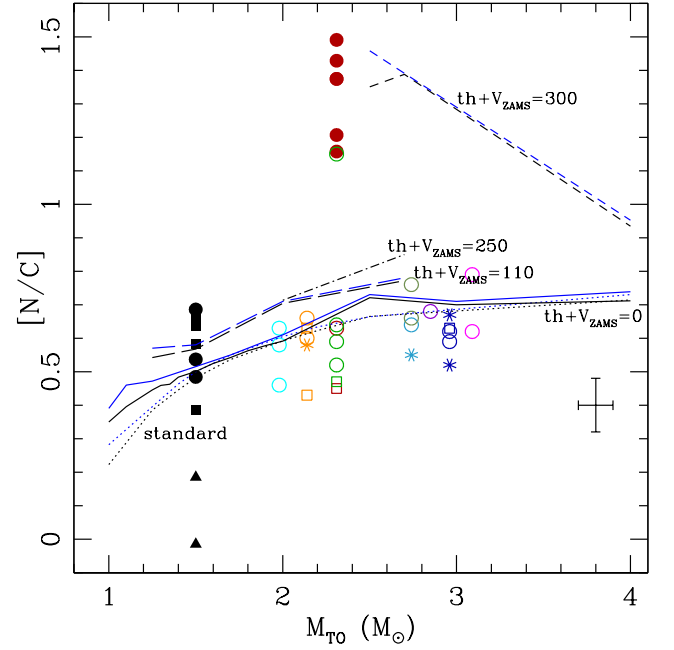}
	  \caption{[N/C] ratio as a function of turn-off mass of the host cluster for the Galactic open cluster sample by \citet {Smiljanic09} and for M67 by \citet {Brown87}.  Symbols and lines have the same meaning as in Fig.\ref{fig:c1213_Mto}.}
	\label{fig:C_N_smiljanic}
\end{figure}

\begin{figure}
	\centering
		\includegraphics[angle=0,width=9cm]{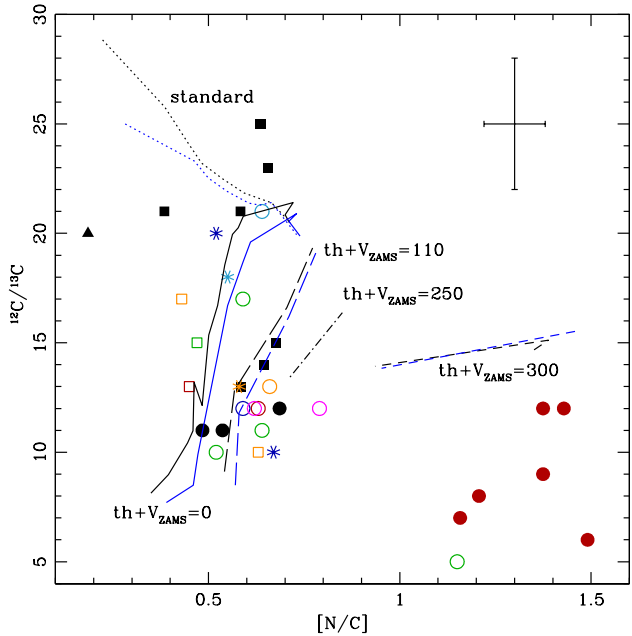}
	  \caption{$^{12}$C/$^{13}$C as a function of [N/C]. See Table~4 for references. Symbols are the same as in Fig \ref{fig:c1213_Mto}.}
	\label{fig:c1213vsCN_smiljanic}
\end{figure}

[N/C] data for the open cluster sample listed in Table 4 is shown as a function of the cluster red turnoff mass and as a function of $^{12}$C/$^{13}$C in Fig.\ref{fig:C_N_smiljanic} and \ref{fig:c1213vsCN_smiljanic} respectively.

During the first dredge-up, the convective envelope of intermediate mass-stars reaches the regions where $^{16}$O has been partially converted into $^{14}$N while the star was on the main sequence, while in the case of low-mass stars it reaches only the first $^{14}$N step due to $^{13}$C-burning (compare the position of the vertical arrows in  Fig.~\ref{fig:profil_1.25_diffrot_Tof}, \ref{fig:profils_abon_rotdiff_2}, and \ref{fig:profils_abon_rotdiff_4}). 
Models thus predict an increase of the post dredge-up [N/C] value with initial stellar mass, in agreement with the observed behaviour. We note, however, that the lower envelope of the observational data lies slightly below the standard predictions, which might indicate that the models overestimate the first dredge-up. On the other hand, the corresponding offset may well be related to observational uncertainties.
Over the whole mass range thermohaline mixing on one hand, and rotation-induced mixing on the other hand, lead to additional  transport of CNO-cycled material, and thus further increase the [N/C] ratio with respect to the standard predictions. Given the observational error bars, one can conclude from Fig.\ref{fig:C_N_smiljanic} and \ref{fig:c1213vsCN_smiljanic} that the models account nicely for the observational constraints on C and N. 

\begin{figure}
	\centering
		\includegraphics[angle=0,width=9cm]{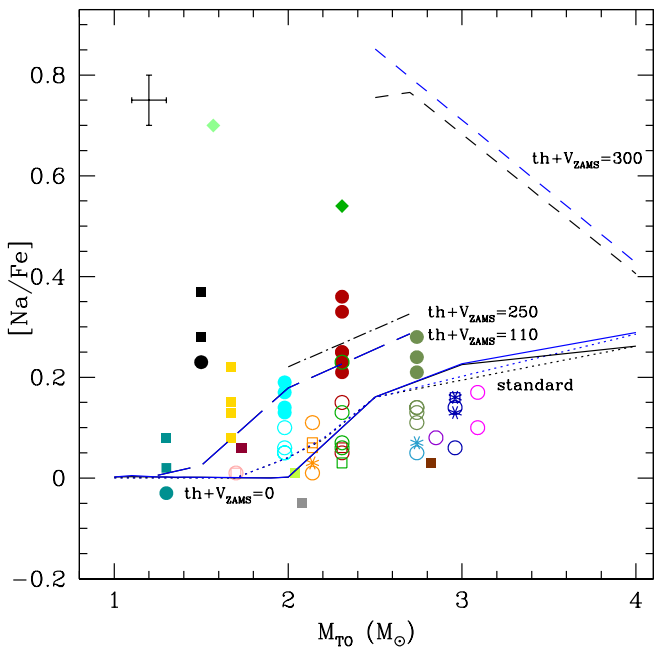}
	  \caption{[Na/Fe] ratio as a function of turn-off mass for the sample of Galactic open clusters listed in Table~4.
	  Symbols and lines are the same as in Fig \ref{fig:c1213_Mto}.
	   The green diamonds are the mean [Na/Fe] values given by \citet{Jacobson07} for IC~4756 and NGC~6939.
	  }
	\label{fig:Na_Fe_smiljanic}
\end{figure}

\begin{figure}
	\centering
		\includegraphics[angle=0,width=9cm]{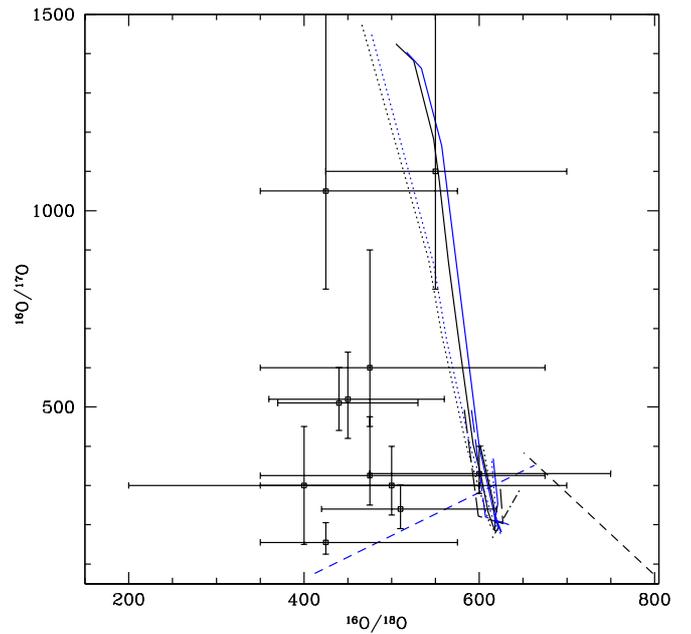}
	  \caption{$^{16}$O/$^{17}$O versus $^{16}$O/$^{18}$O. Observations are from \citet{HaLa84}, \citet{HaLa88} for stars in the 1--3M~$_{\odot}$ range. The line symbols are the same as in Fig.\ref{fig:c1213_Mto}. The initial values assumed for $^{16}$O/$^{17}$O  and $^{16}$O/$^{18}$O are 1490 and 445 respectively.
	  }
	\label{fig:oxygenHarris}
\end{figure}

In figure \ref{fig:Na_Fe_smiljanic} we plot the [Na/Fe] ratio for the open cluster sample listed in Table 4 as a function of cluster turnoff mass.
Note that the observational data were reported to the solar Na value from \citet{Asplund05} we assume in the initial composition of our 
models\footnote{\citet{Smiljanic09} and \citet{Mikolaitisetal10} adopted solar abundances recommended by \citet{GreSau93} (A(Na) = 6.33, A(Fe)=7.50), while \citet {Jacobson07, Jacobson08, Jacobson09} and \citet{Hamdani00} used solar abundances given by \citet{AndGre89} (A(Na)=6.33). In Fig. \ref{fig:Na_Fe_smiljanic}  the observational data are reported to the solar abondances values we use in our computations \citep[A(Na)=6.20, and A(Fe) = 7.45]{Asplund05}.}.
Both the predictions and observations show a positive correlation between  [Na/Fe] values and stellar mass. Rotation-induced mixing leads to an increase of the amount of Na processed to the surface and allows an explanation for the observed dispersion. There is however an offset of about 0.1~dex between the data and the predictions as was already noticed by \citet{Smiljanic09} who compared their observations with standard model predictions by \citet{Mowlavi99}.  
As a matter of fact, very different observational Na abundances for giant stars have been reported in the literature. As can be seen in Fig.~\ref{fig:Na_Fe_smiljanic} some studies present [Na/Fe] values as high as +0.6~dex \citep{Jacobson07}, 
some only a mild overabundance of +0.2~dex \citep [see Fig.~\ref{fig:Na_Fe_smiljanic}] {Hamdani00}, and other solar values \citep{Sestito07}. We refer to \citet{Smiljanic09} for a discussion on the possible causes of these discrepancies. 
The present predictions for the stars more massive than $\sim$ 2~M$_{\odot}$ are actually in better agreement with the mild overabundance of [Na/Fe] measured by \citet{Hamdani00}.

Finally we show in Fig.~\ref{fig:oxygenHarris} the $^{16}$O/$^{17}$O vs $^{16}$O/$^{18}$O for the G and K giants by \citet{HaLa84} and \citet{HaLa88}. Included in the figure are our predictions. As discussed in \S~3, thermohaline mixing affects only slightly the $^{16}$O/$^{18}$O ratio, and leaves $^{16}$O/$^{17}$O unaffected; on the other hand, rotation-induced mixing lowers the $^{16}$O/$^{17}$O ratio, and helps account for the lowest $^{16}$O/$^{18}$O values. Given the large observational uncertainties, the predictions are reasonably consistent with the O isotopic ratios measured in RGB stars.

%============================================================= Conclusions =======================================================

\section{Conclusions}

In the present paper we have investigated the effects of the thermohaline instability induced by $^3$He-burning that sets in above the RGB bump and of rotation-induced mixing on the evolution and chemical properties of low- and intermediate-mass stars (1 to 4~M$_{\odot}$) at solar metallicity. 
All the stellar models were computed up to the end of the second dredge-up on the early-AGB, and some of them up to the end of the TP-AGB phase. 
Predictions are compared to data for lithium, $^{12}$C/$^{13}$C, [N/C], [Na/Fe], $^{16}$O/$^{17}$O, and $^{16}$O/$^{18}$O in giant stars with well-defined masses and evolutionary status on the RGB, clump, early-AGB, and planetary nebulae phases. 

We find that the theoretical and observational behaviours for these species are in very good agreement over the whole scrutinized mass range. 
Thermohaline mixing is confirmed to be the main physical process governing the surface abundances of $^3$He, $^7$Li, C, and N for stars more evolved than the RGB bump in all the models with initial masses below 2.2~M$_{\odot}$, although its efficiency is 
increasing with decreasing initial stellar mass. 
In all cases $^3$He decreases by a large fraction in the stellar yields compared to the standard models, although we find that low-mass stars remain net producers of $^3$He ($^3$He yields for stellar models over a broad range in both mass and metallicity will be published in a future paper).
It is also found that thermohaline mixing leads to lithium production on the TP-AGB phase, as first shown by \citet{Stancliffe10} in the case of low-metallicity stars. However, the Li yields remain negative, and these stars are not expected to contribute to Galactic Li enrichment.
In one 2.0~M$_{\odot}$ model computed up to the AGB tip thermohaline mixing was found to help initiating the occurrence of the third dredge-up during the TP-AGB phase.

On the other hand, rotation-induced mixing modifies the internal chemical structure of main sequence stars, although its signatures are revealed only later in the evolution when the first dredge-up occurs. It favours the occurrence of thermohaline mixing in RGB stars in the mass range between $\sim$ 1.5 and 2.2~M$_{\odot}$. 
It accounts for the observed dispersion of abundances in stars of similar mass and evolutionary status, and is necessary to explain the features of CN-processed material in intermediate-mass stars.

These results were obtained using the prescription for the turbulent diffusivity related to the thermohaline instability advocated first by \citet{Ulrich72} that is supported by laboratory experiments \citet{Krish03}. The same prescription was shown by \citet {ChaZah07a} to nicely account for the photospheric composition of low-mass, low-metallicity giant stars. 
 
\begin{acknowledgements}
We wish to thank Jean-Paul Zahn and Thibaut Decressin 
%as well as the anonymous referee 
for helpful comments on our manuscript.
We acknowledge financial support from the Swiss National Science Foundation (FNS) and the french Programme National de Physique Stellaire (PNPS) of CNRS/INSU.
This research has made use of the VizieR catalogue access tool, CDS, Strasbourg, France.
\end{acknowledgements}

\begin{appendix}
\section{Nuclear reaction rates}

All reactions for hydrogen burning are computed with nominal NACRE reaction rates \citep{Argulo99}, with the exception of : 

$^{14}$C(p,$\gamma)^{15}$N \citep{Wiescheretal90}; 
$^{14}$C(p,n)$^{14}$N \citep{KoeObr89};
$^{14}$C(p,$\alpha)^{11}$B \citep{CauFow88};
$^{14}$N(p,$\gamma)^{15}$O \citep{Mukhamedzhanov03};
$^{21}$Ne(p,$\gamma)^{22}$Na, 
$^{22}$Na(p,$\gamma)^{23}$Na, 
$^{23}$Na(p,$\alpha)^{20}$Ne, 
and $^{23}$Na(p,$\gamma)^{24}$Mg \citep{Illiadisetal01}; 
$^{22}$Ne(p,$\gamma)^{23}$Na \citep{Haleetal02}.

\end{appendix}

\bibliographystyle{aa}
\bibliography{Reference}

\end{document}